%% file: sample-authordraft.tex
\documentclass[sigconf]{acmart}
\AtBeginDocument{%
  \providecommand\BibTeX{{%
    \normalfont B\kern-0.5em{\scshape i\kern-0.25em b}\kern-0.8em\TeX}}}
\long\def\comment#1{}

\usepackage{
  tikz,
  pgfplots,
  pgfplotstable
}
\usepackage{hyperref}
\usepackage{pifont}
\usepackage{algorithm}
\usepackage{algorithmic}
\usepackage{enumitem}

\newcommand{\cmark}{\text{\ding{51}}}
\newcommand{\xmark}{\text{\ding{55}}}
\newcommand{\makecell}{}

\usetikzlibrary{
  shapes.geometric,
  arrows,
  external,
  pgfplots.groupplots,
  matrix
}

\begin{document}

\title{Generating Practical Adversarial Network \\ Traffic Flows Using NIDSGAN}

\author{Bolor-Erdene Zolbayar}
\affiliation{%
    \institution{The Pennslyvania State University}
    \country{United States}
}
\email{bzz5065@psu.edu}
\author{Ryan Sheatsley}
\affiliation{%
    \institution{The Pennslyvania State University}
    \country{United States}
}
\email{sheatsley@psu.edu}
\author{Patrick McDaniel}
\affiliation{%
    \institution{The Pennslyvania State University}
    \country{United States}
}
\email{mcdaniel@cse.psu.edu}
\author{Michael J. Weisman}
\affiliation{%
    \institution{Army Research Laboratory}
    \country{United States}
}
\email{michael.j.weisman2.civ@army.mil}
\author{Sencun Zhu}
\affiliation{%
    \institution{The Pennslyvania State University}
    \country{United States}
}
\email{sxz16@psu.edu}
\author{Shitong Zhu}
\affiliation{%
    \institution{UC Riverside}
    \country{United States}
}
\email{shitong.zhu@email.ucr.edu}
\author{Srikanth Krishnamurthy}
\affiliation{%
    \institution{UC Riverside}
    \country{United States}
}
\email{krish@cs.ucr.edu}
\renewcommand{\shortauthors}{Zolbayar et al.}

\input{1.Abstract}
{}


%


\maketitle
\input{2.Introduction}

\input{3.Background}
\input{4.Approach}
\input{5.Evaluation}

\input{6.Discussion}

\input{7.Conclusion}


\bibliographystyle{ACM-Reference-Format}
\bibliography{bib}


\input{appendix}

\end{document}

%% file: 1.Abstract.tex
\begin{abstract}
Network intrusion detection systems (NIDS) are an essential defense for computer networks and the hosts within them. Machine learning (ML) nowadays predominantly serves as the basis for NIDS decision making, where models are tuned to reduce false alarms, increase detection rates, and detect known and unknown attacks. At the same time, ML models have been found to be vulnerable to adversarial examples that undermine the downstream task. In this work, we ask the practical question of whether real-world ML-based NIDS can be circumvented by crafted adversarial flows, and if so, how can they be created. We develop the generative adversarial network (GAN)-based attack algorithm NIDSGAN and evaluate its effectiveness against realistic ML-based NIDS. Two main challenges arise for generating adversarial network traffic flows: (1) the network features must obey the constraints of the domain (i.e., represent realistic network behavior), and (2) the adversary must learn the decision behavior of the target NIDS without knowing its model internals (e.g., architecture and meta-parameters) and training data. Despite these challenges, the NIDSGAN algorithm generates highly realistic adversarial traffic flows that evade ML-based NIDS.  We evaluate our attack algorithm against two state-of-the-art DNN-based NIDS
in whitebox, blackbox, and restricted-blackbox threat models and achieve success rates which are on average 99\%, 85\%, and 70\%, respectively. We also show that our attack algorithm can evade NIDS based on classical ML models including logistic regression, SVM, decision trees and KNNs, with a success rate of 70\% on average. Our results demonstrate that deploying ML-based NIDS without careful defensive strategies against adversarial flows may (and arguably likely will) lead to future compromises.


\end{abstract}

%% file: 2.Introduction.tex
\section{Introduction} 


NIDS are the {\it de facto} standard for detecting malicious activities in computer networks~\cite{mukherjee1994network, axelsson2000intrusion}. They can foil existing malware attacks such as backdoors, trojans, and rootkits~\cite{malware1,malware2} and detect social engineering attacks such as phishing and man-in-the-middle attacks ~\cite{phishing1, phishing2}. Most frequently ML techniques play an important role in NIDS architectures because they provide lower false alarm rates, higher detection rates, and better capabilities of finding mutations of known and unknown attacks not detectable by conventional techniques ~\cite{mlbetter_0, mlbetter_1, alertnet, mlbetter_3}. 

However, ML models are vulnerable to adversarial examples~\cite{adversarialExamples1,adversarialExamples2,adversarialExamples3,adversarialExamples4}. In this work, we posit that deploying ML-based NIDS without analyzing the possible risks of adversarial traffic in these safety-critical systems can be catastrophic. 
To explore the generation of adversarial flows, one has to consider two challenges. First, flow feature values are dictated by rules (constraints) determined by network behavior~\cite{rs}. While domains such as images allow the adversary to perturb any feature by any amount, highly constrained domains such as networking have feature and intra-feature constraints that vastly shrink the ``legal'' perturbation space. Any created flow that violates the feature constraints will result in an illegal or unrealizable behavior (and is thus manifestly adversarial to the NIDS). Second, in realistic environments, the adversary can not have direct access to the NIDS' model and its training data. Therefore, the adversary must train a local model using data collected by observing NIDS behavior (which is likely to be vanishingly small in comparison to the training data). Note that the efficacy of adversarially generated flows to bypass NIDS remains an open question. Prior efforts at crafting adversarial sample flows to bypass or otherwise manipulate NIDS~\cite{back1, malgan, nattack, idsgan} ignore (or use extremely weak) constraints and assume substantial access to a model, architecture, and/or training data. Discussed in the following section, these prior efforts--while technically sophisticated attacks on ML--do not yield attacks that were grounded in realistic networks and threat models.

In this work, we develop a GAN-based attack algorithm NIDSGAN that generates realistic adversarial network traffic flows. These flows evade ML-based NIDS in whitebox, blackbox, and restricted-blackbox threat models.  We choose GANs for the basis of our attack algorithm because GANs are (a) self reinforcing as they learn the distribution of the generating class and (b) known to be exceptional at generating realistic adversarial samples in complex domains, e.g., image, speech, and text. We find that the adversary can have the ability to observe the target NIDS' response to a very small number of adversarial flows (i.e., perform limited reconnaissance in the target network). Given this, the adversary attempts to add a minimal perturbation to the feature values of a real attack flow to change the NIDS’ prediction from ``an attack class" to ``a benign class". We consider {\it whitebox} settings where an adversary has full knowledge of the target models' training data, architecture, and parameters and {\it blackbox} settings where an adversary has no knowledge about the model internals and limited training data of the target NIDS. Lastly, we consider a {\it restricted-blackbox} threat model that represents a realistic environment where the number of samples available to the adversary is strictly limited. In this latter case we apply active learning~\cite{activeLearning} to amplify the training data and ultimately improve the success rate of the attack. Although the active learning-based techniques have been proposed in the previous work of adversarial machine learning (AML)~\cite{papernot2016practical, activeIlyas}, they have not been studied in the constrained domains.   In the context of network intrusion detection, NIDSGAN has three advantages over a naive application of AML algorithms: (1) it generates highly realistic adversarial traffic flows in realistic threat models, (2) it enforces the network domain constraints, and  (3) the adversary remains successful in the presence of very limited training samples.

In the blackbox threat models, we evaluate against two NIDS models proposed in the literature~\cite{alertnet, deepnet} as well as an exemplar trained model. We train the models on the NSL-KDD and CICIDS-2017 datasets. In the restricted blackbox threat model, we consider two restrictions for the attack: (1) we limit the size of adversary's training data (less than 1\% of the original datasets) (2) we restrict the number of attack traffic flows launched to no larger than three times the size of the local model training set in the active learning cases. Our experiments demonstrate that NIDSGAN is an effective framework for evading ML-based NIDS in all three whitebox, blackbox, and restricted-blackbox setups. In the whitebox, the adversary can bypass any model of NIDS with a success rate of nearly 100\%. In the blackbox threat model, NIDSGAN achieves a success rate of over 90\% against NIDS models trained with CICIDS-2017, which is exceptional as the adversary does not have access to the training data and model internals of the target model. However, NIDSGAN without active learning has a less than 70\% success rate against NIDS models trained with NSL-KDD in the restricted-blackbox setup (with <740 samples).  We show that we can improve our success rates by as much as 22\% using active learning. In this work, we make the following contributions:
\begin{enumerate}
    \item We develop NIDSGAN that generates adversarial network traffic flows that evade ML-based NIDS. \vspace{6pt}
    
    \item We evaluate NIDSGAN against two real-world NIDS models from the literature as well as an exemplar model.  We observe success rates greater than 85\% and 70\% on blackbox and restricted-blackbox threat models, and nearly 100\% in the whitebox setup. \vspace{6pt}
    
    \item We propose an active learning approach that helps to (a) increase the attack success rates by up to 22\%. (b) reduce the training samples by three times but still achieve the same levels of success rates of the cases without active learning. \vspace{6pt}
    
    \item We compare NIDSGAN against two existing GAN-based attack frameworks in a restricted-blackbox threat model. NIDSGAN (75\%) outperforms the ADVGAN (0.5\%) and IDSGAN (19\%) achieving more than 56\% higher success rates than the other two attacks. 
 \end{enumerate}

%% file: 3.Background.tex
\section{Background}

\subsection{Network Intrusion Detection Systems}
Misuse-based NIDS are divided into two main categories: knowledge-based and ML-based. In knowledge-based NIDS (also known as a rule-based detection system), attack network traffic flows are directly compared with previously defined rules or attack patterns. One of the drawbacks of rule-based NIDS is that they are not able to detect mutations of attacks~\cite{mishra2019}, specifically in the dynamic environments of modern computer networks where the attacks are mutating unexpectedly. ML-based misuse detection systems, on the other hand, learn from signatures in the dataset and predict the possible mutations of known attacks~\cite{mishra2019,generalKnowledge1}. In this work, we will specifically focus on ML-based misuse NIDS with multiple-class detection capabilities.

\subsection{Adversarial Examples}

Deep neural networks (DNNs) can easily be manipulated by adversarial examples. In 2013, Szegedy et al.~\cite{adversarialExamples1}, and Biggio et al.~\cite{adversarialExamples2}, first discovered the existence of adversarial examples in the image domain. By applying a small perturbation unnoticeable to human eyes to an original instance of an image, it is able to change the classification of a trained model for the image. By exploiting this vulnerability of neural networks, adversaries can potentially manipulate self-driving cars, smart devices, and intrusion detection systems that rely on DNNs. Given an input $\vec{x}$ and a trained target model $F(\vec{x})$, an adversary tries to find the minimum perturbation ${{\Delta \vec{x}}}$ under some norm (commonly $l^0, l^2, \textrm{or } l^\infty$)~\cite{adversarialExamples3} to cause the $F(\vec{x}+ \Delta \vec{x})$ to be classified as a specific target label $t \neq F(\vec{x})$~\cite{adversarialExamples4}. In Equation \ref{eqn:1}\cite{adversarialExamples3}, this can generally be expressed by an optimization problem where $c$ is a constant that is found by binary search and $t$ is the target class~\cite{adversarialExamples3,adversarialExamples4} and ``$\textrm{Loss()}$'' is the cross-entropy loss.
    \begin{equation} 
    \label{eqn:1}
    \begin{aligned}
    & \min_{\Delta \vec{x}} c|\Delta \vec{x}| + \textrm{Loss}(F(\vec{x}+\Delta \vec{x}),t) \\
    & \text{such that } (\vec{x} + \Delta \vec{x}) \in [0,1]^n \\
    \end{aligned}
    \end{equation}

\noindent where $n$ is the dimension of the vector $\vec{x}$. As mentioned, these works have focused on unconstrained domains, specifically in the image domain. However, most of the domains in the real world are constrained. In this work, we create adversarial examples in a network intrusion detection domain where we determine and enforce the domain constraints. In our problem, $F$ is the target ML-based NIDS we want to fool and the target class $t$ is the benign class. 

The discovery of adversarial examples has inspired researchers to develop algorithms of greater and greater sophistication, both for attacking and defending against adversarial attacks. In 2014, Goodfellow et al. found an efficient algorithm that maximizes the loss function by adding a perturbation $\Delta x$ bounded by the
$l^\infty$ norm and $\Delta x_i \in [-\epsilon, \epsilon]$ in the direction of the gradient as shown in the Equation~\ref{equation:goodfellow}.

    \begin{equation} 
    \begin{aligned}
    \label{equation:goodfellow}
    &  \Delta x =  \nabla_x [loss(F(x+\Delta x),t)] \\
    & \text{such that } x' = x + \Delta x \in [0,1]^n
    \end{aligned}
    \end{equation}

In 2016, Papernot et al. first introduced JSMA that minimizes over number of features changed by the perturbation. In the following year, Papernot et al. also introduced a black-box attack which represents a practical attack scenario where an adversary does not have any knowledge of target model internals and its training data. Before this, all attacks considered whitebox attacks where the adversary is accessible to the target model internals or its training data. Carlini and Wagner aimed to minimize the perturbation under different norms with different objective functions. 
    \begin{equation} 
    \begin{aligned}
    \label{equation:cw}
    & \min [(\max_{i\neq t} F(x+\Delta x)_i) - F(x+\Delta x)_t]
    \end{aligned}
    \end{equation}

They argue that the perturbation under $l^2$ norm $\|\Delta x\|_2$ is almost imperceptible to human eyes and introduced adversarial objective functions including the one in Equation~\ref{equation:cw}. The implication of this function on an attack traffic flow is to increase the confidence score of benign class while decreasing the confidence scores of the other classes in its classification by the NIDS. After Carlini and Wagner attack, in 2017, Madry et al. introduced a PGD whitebox attack~\cite{madry2017towards} that finds adversarial examples by finding a perturbation that maximizes the model loss on a specific input bounded by 0.3 $l^\infty$ norm ball. 

\subsection{Generative Adversarial Networks}
GANs~\cite{gan1}, generative models first created in 2014, have achieved unprecedented success in computer vision and natural language processing. GANs consist of two competing neural networks: a generator and a discriminator.

    \begin{figure}[h]
    \centering
    \includegraphics[width=0.48\textwidth]{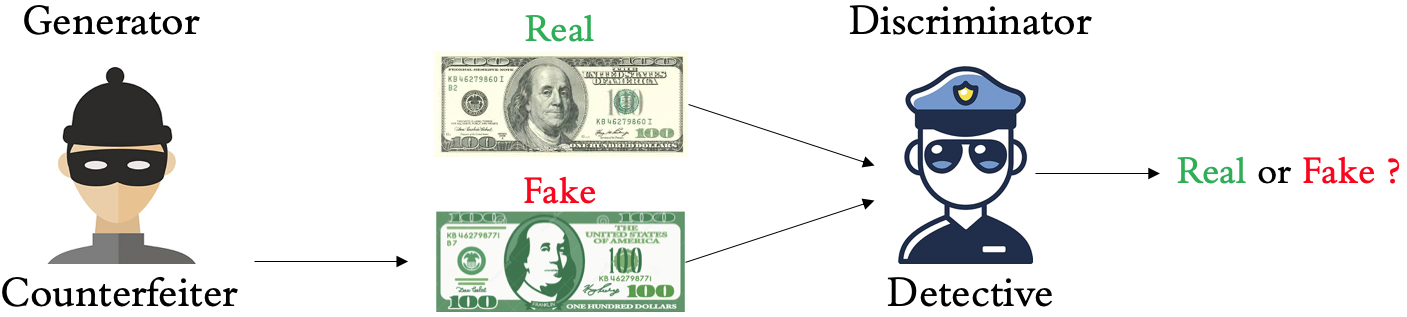}
    \caption{Generative Adversarial Networks}
    \label{fig:howGans}
    \end{figure}

We can think of the generator as a counterfeiter who is trying to generate a fake currency; on the other hand, the discriminator is a detective who is trying to correctly predict if the given currency is real or fake. The competition between these two neural networks improves each other's performance and eventually, the generator outputs examples that are indistinguishable from the examples of the training data. We use GANs in our algorithm for two reasons: 1) GANs functionality in our objective function will strive to optimize their loss functions such that the generated adversarial examples are from the set of the original attack traffic. 2) Other GAN-based attack algorithms such as AdvGAN~\cite{advgan} are known to be highly effective at fooling DNN models in other domains such as images~\cite{advgan}.

\subsection{Crafting Adversarial Examples in the Network Domain}
Recently, some works ~\cite{back1, malgan, nattack, idsgan} have studied evading ML-based NIDS in the feature space. Evading ML-based NIDS requires defining and enforcing the constraints of the network domain on the adversarial traffic flows without losing their semantics. However, most of the previous works for evading NIDS do not consider the constraints in their process of creating adversarial examples~\cite{back1,nattack}. Some works~\cite{malgan,idsgan} define and enforce the domain constraints, but, they consider simplistic assumptions on the threat model and constraints. Therefore, they do not have restrictions on the amount of perturbation on the features of attack traffic, which can break the semantics of the attack traffic flows. In general, previous works have the following flaws: a) Previous works do not consider valid or any domain constraints of network intrusion detection systems for crafting their adversarial traffic flows. b) They assume that the adversary can perturb a set of features by any amount, which can break the semantics of the attacks. c) Most of the work has been done in whitebox settings by assuming that the adversary is an insider, which is not always the case in the practical scenarios; the rest of the work done in blackbox settings does not have valid assumptions in their threat model. 

\subsection{Crafting Adversarial Examples in the Network Domain using GANs}
\begin{figure}[h]
    \centering
    \includegraphics[width=0.48\textwidth]{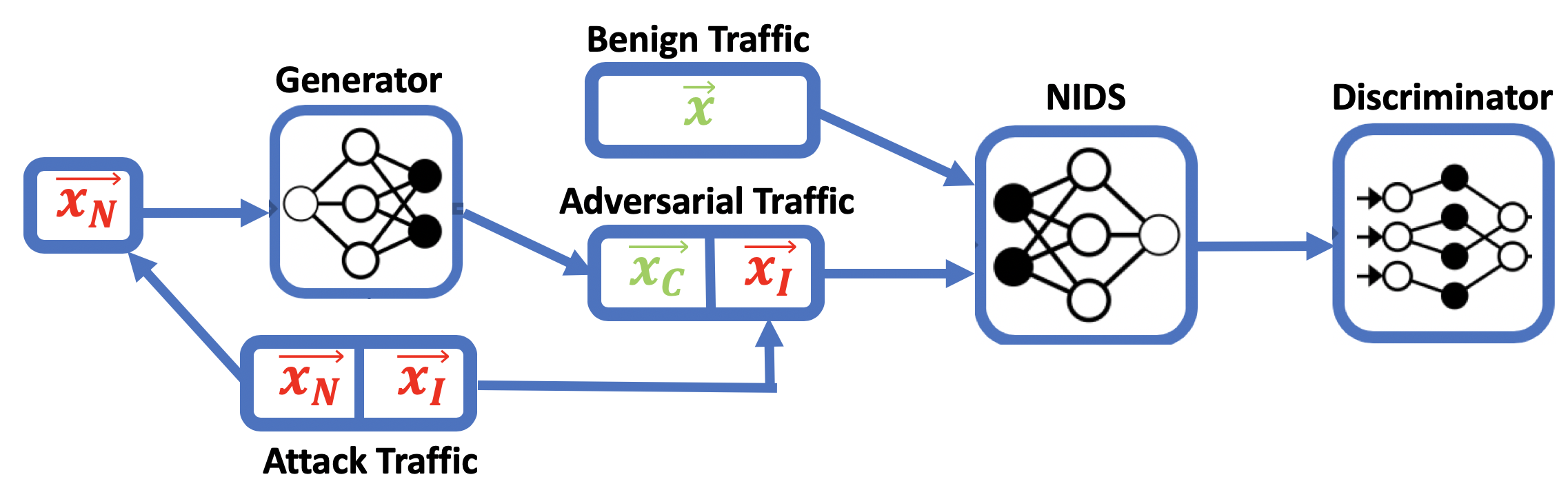}
    \caption{IDSGAN}
    \label{fig:idsgan}
    \end{figure}

There have been three GAN-based algorithms introduced in the past for crafting adversarial examples in the network domain: Gen-AAL~\cite{tucker_gan}, IDSGAN~\cite{idsgan}, and NGAN~\cite{idsgan_copy}. The downside of Gen-AAL algorithm is that it does not take the domain constraints into account for generating the adversarial traffic flows. In other words, the generated traffic flows will not function in the computer networks because they break the domain constraints. NGAN is a duplication work of the IDSGAN in a different setup with a different dataset. IDSGAN, depicted in Figure~\ref{fig:idsgan}, first divides features of attack traffic into two groups, important features $\vec{x}_I$ that preserve the semantics of the attack traffic and unimportant features $\vec{x}_N$ that do not have any effect on the semantics of the attack traffic. Then, the IDSGAN's generator takes the unimportant features $\vec{x}_N$ as an input and replaces them with $\vec{x}_C$ without any constraints enforced. This algorithm has two main downsides: (1) No restriction on the perturbation in their algorithm can easily cause to generate invalid network traffic flows, which is proved in the compared results in the evaluation section. (2) The choice of a set of ``fixed'' important features is based on prior and marginally related work. This is insufficient because categorizing attack traffic's features into important and unimportant groups is inherently complex without a proven instance and domain expertise. In essence, the important features of specific attacks can be dynamic varying from one computer network to a different computer network. 

In their threat model, the IDSGAN uses half of the NSL-KDD's training dataset for training the NIDS model and the other half for training the attack algorithm. This has two downsides: (1) Cutting the training dataset of NIDS in half can reduce the detection capability of the NIDS significantly; therefore, the evaluation of the attack algorithm against the NIDS model becomes biased. (2) The threat model assumes that the adversary has the half of the training data for training the attack algorithm. This threat model can be considered as a whitebox threat model where the adversary has the whole training data of the NIDS model. However, in most cases of practical scenarios, the adversary can not know the model internals and the training data of the NIDS.



    



In this work, by iteratively analyzing the response of the target NIDS for the adversarial examples on the fly, we use an active learning technique to gain additional information about the target model's decision boundary and train the local model more efficiently. With a local model trained with a sufficient number of adversarial examples, eventually, the adversarial examples that bypass the local model will most likely bypass the target NIDS model. In this work, we have two objectives for the improvement of the blackbox attack: (1) reduce the initial training size of the local model; (2)  significantly improve the attack success rate with active learning by systematically deploying adversarial examples and observing their classification labels.

%% file: 4.Approach.tex
\section{Approach}
In this section, we detail our threat model, describe the objectives of our attack algorithm, discuss how to integrate the domain constraints in the attack, and present the active learning technique for (1) effectively reducing the required training dataset size or (2) increasing the success rate with the same training dataset.


\begin{figure*}[t]
    \centering
    \includegraphics[width=0.70\textwidth]{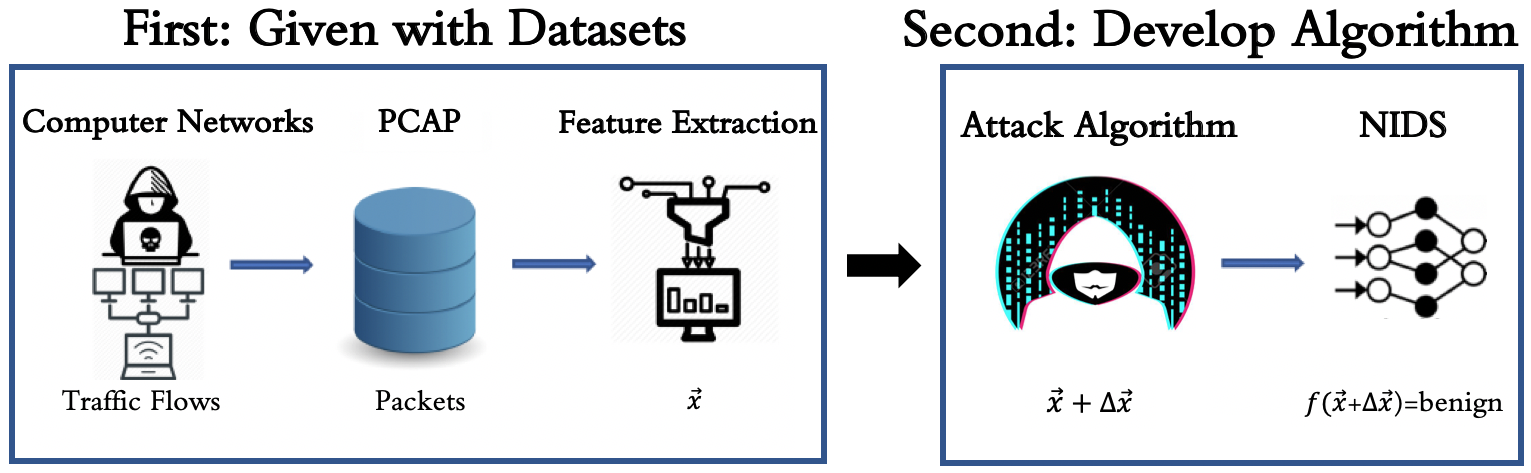}
    \caption{The steps described in the first box are done by the dataset providers. First, network traffic flows are simulated on computer networks and stored into PCAP format. Afterward, traffic flows are extracted from the PCAP data using CICFlowMeter. The second box shows the scope of this work. We are developing an attack algorithm against ML-based NIDS models trained with given network datasets.}
    \label{fig:FlowChart}
\end{figure*}

The attack, conceptually, is shown in Fig. \ref{fig:FlowChart}. In this work, we solely focus on the feature space of the network traffic after the feature extraction step, as shown in Fig. \ref{fig:FlowChart}. The attacks are first simulated on real computer networks, and then captured as PCAP files to form the datasets. Both datasets CICIDS-2017~\cite{cicids-2017} and NSL-KDD~\cite{nsl-kdd1} provide intrinsic, content, host-based, and time-based features of network flow extracted from the PCAPs data using programs such as CICFlowMeter~\cite{CICFlowMeter}. With these features, three DNN-based NIDS models are trained with the given datasets. Afterward, adversarial traffic flows are created under whitebox, blackbox, as well as restricted-blackbox threat models using our GAN-based attack algorithm.  


\definecolor{green1}{rgb}{0., 0.7, 0.}
\definecolor{red1}{rgb}{0.9, 0., 0.}
\newrobustcmd*{\newsquare}[1]{\tikz{\filldraw[draw=#1,fill=#1] (0,0) rectangle (0.35cm,0.35cm);}}
\newrobustcmd*{\filledcircle}[1]{\tikz{\filldraw[draw=#1,fill=#1] (0,0) circle [radius=0.175cm];}}
\newrobustcmd*{\filledtriangle}[1]{\tikz{\filldraw[draw=#1,fill=#1] (0,0) --(0.4cm,0) -- (0.2cm,0.4cm);}}
\newrobustcmd*{\shape}[1]{\newsquare{#1}}
\newrobustcmd*{\ttrue}{\newsquare{green1}}
\newrobustcmd*{\ffalse}{\filledcircle{red1}}


\subsection{Threat Model}
{}

In this work, we evaluate our attack algorithm in three threat models: whitebox, blackbox, and restricted-blackbox. 
The adversary performs an evasion attack by forcing the NIDS to misclassify attack network traffic flow as benign. The first set of experiments are done in whitebox threat model where an adversary can access all training samples used in the NIDS model, which is an ideal (but likely unrealistic) scenario for the adversary. The second set of experiments will be in blackbox threat model where the adversary has no knowledge of NIDS's architectural information and training data, which is a more constrained threat model for the adversary.

\begin{figure}[h]
    \centering
    \includegraphics[width=0.4\textwidth]{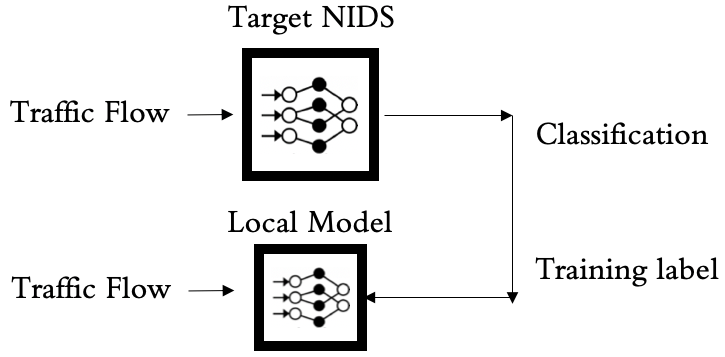}
    \caption{Adversary is training a local model in the blackbox and the restricted-blackbox threat models}
    \label{fig:blackbox}
\end{figure}

In this scenario, to gain information about the NIDS, the adversary trains a local model by launching network traffic flows in the networks (which are separate from the NIDS's training data) and producing labels for the launched network flows by analyzing the response of the NIDS. By training the local model with the feedback of the target NIDS, the local model can imitate the decision behavior the target NIDS model, and adversarial examples generated to bypass the local model can bypass the target NIDS. The adversary can find out the NIDS' prediction attack classes and feature extraction tool as the most network traffic flow extraction tools are common and available. To define the threat model characteristics of the blackbox settings more in detail, we list the objectives, capabilities, and strategies of the adversary in Table.~\ref{tab:threat_model}

\begin{table}[t]
\centering
\caption{Adversary's Strategies and Capabilities} \label{tab:threat_model}
\scalebox{0.8}{
\begin{tabular}{|p{17mm} | p{35mm} | p{12mm} | p{12mm} | p{13mm}|} 
\hline
Attacker's & Definitions/Threat Models & Whitebox & Blackbox & Restricted-blackbox\\ 
\hline
Objective & \makecell{\hspace*{0.6cm} Evasion attack }  & \makecell{\hspace*{0.5cm}\cmark } & 
\makecell{   \hspace*{0.5cm} \cmark } & 
\makecell{\hspace*{0.5cm} \cmark   } \\
 \hline
Access \& Capability & 
\makecell{
\hspace*{0.1cm} NIDS training algorithm \newline \hspace*{0.2cm} NIDS training data \newline \hspace*{0.4cm} NIDS test data \newline \hspace*{0.15cm} Small set of traffic flows \newline \hspace*{0.2cm} Identify classification \newline \hspace*{0.15cm} Feature extraction tool}  

 & \makecell
 
 {\hspace*{0.4cm} \xmark \newline \hspace*{0.4cm} \cmark \newline \hspace*{0.4cm} \xmark \newline \hspace*{0.4cm} \xmark  \newline \hspace*{0.4cm} \cmark \newline \hspace*{0.4cm} \cmark  } 
 
 & \makecell
 
 {\hspace*{0.5cm} \xmark \newline \hspace*{0.5cm} \xmark \newline \hspace*{0.5cm} \xmark \newline \hspace*{0.5cm} \cmark  \newline \hspace*{0.5cm} \cmark \newline \hspace*{0.5cm} \cmark }
 
 & \makecell
 
 {\hspace*{0.5cm} \xmark \newline \hspace*{0.5cm} \xmark \newline \hspace*{0.5cm} \xmark \newline \hspace*{0.5cm} \cmark    \newline \hspace*{0.5cm} \cmark \newline \hspace*{0.5cm} \cmark } \\
 \hline
 \hspace*{0.05cm} 
 Strategy & \makecell{\hspace*{0.4cm} Train local model \newline \hspace*{0.1cm} Apply active learning \newline \hspace*{0.3cm} Enforce constraints}  
 & \makecell{\hspace*{0.4cm} \xmark \newline \hspace*{0.4cm} \xmark \newline \hspace*{0.4cm} \cmark } 
 & \makecell{ \hspace*{0.5cm} \cmark \newline \hspace*{0.5cm} \xmark \newline \hspace*{0.5cm} \cmark }
 & \makecell{ \hspace*{0.5cm} \cmark \newline \hspace*{0.5cm} \cmark  \newline \hspace*{0.5cm} \cmark }\\
 \hline
\end{tabular}
}
\end{table}

\subsection{NIDSGAN Algorithm}

\begin{figure}[h]
    \centering
    \includegraphics[width=0.45\textwidth]{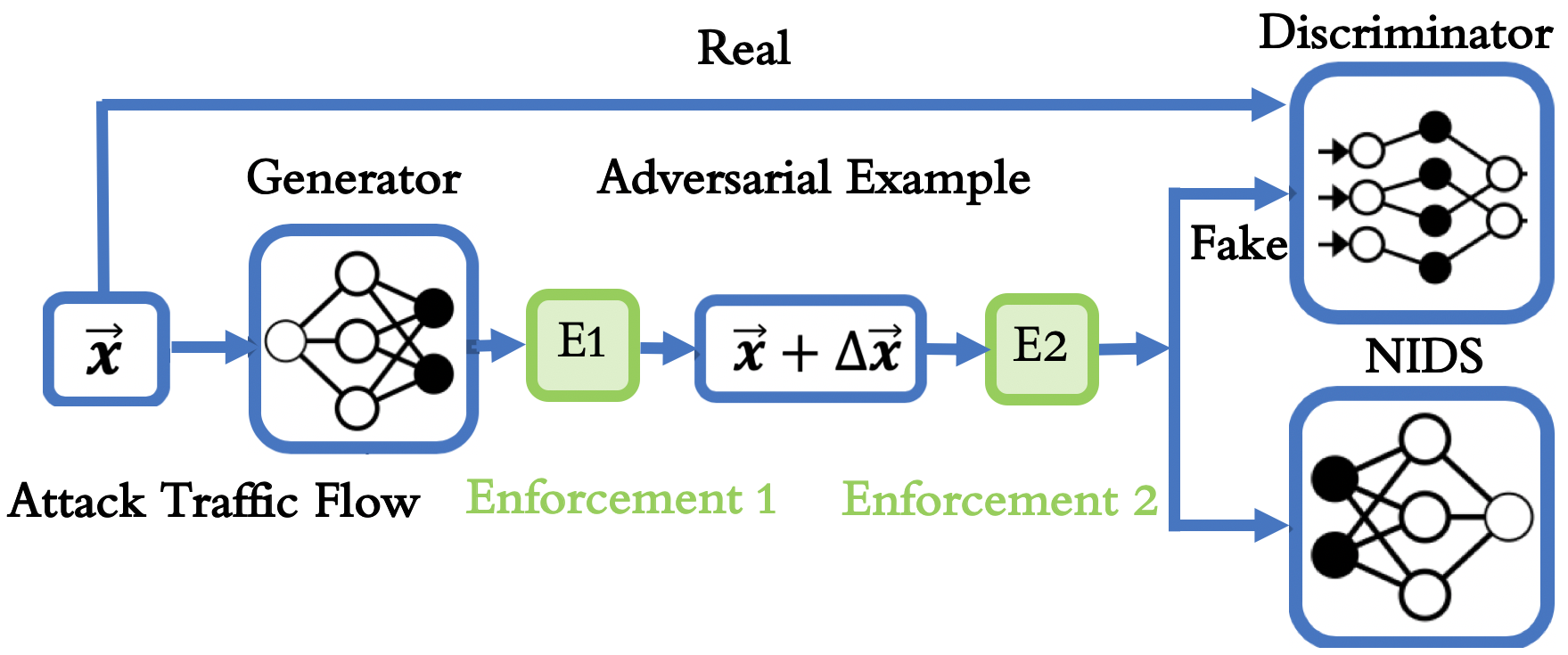}
    \caption{NIDSGAN Structure}
    \label{fig:attack}
\end{figure}

The main objective of our algorithm is to add a minimum possible perturbation $\Delta x$ to features of attack traffic $x$ such that the resulting adversarial traffic $x + \Delta x$ successfully bypasses DNN-based NIDS while maintaining its domain constraints and attack semantics as shown in Figure~\ref{fig:attack}. To generate such adversarial traffic flows, we train our attack algorithms' generator for three objectives by optimizing the corresponding terms in the loss function: (1) $L_\textrm{adversarial}$ makes sure that the resulting adversarial traffic flow is classified as benign by the target NIDS. (2) $L_\textrm{perturbation}$ optimizes to minimize the perturbation applied to the adversarial traffic flow. (3) $L_\textrm{wgan-gp}$ makes sure that the resulting adversarial traffic flow is indistinguishable from the real attack traffic flow. These three loss terms are considered jointly in the generator, as defined in Equation~\ref{eqn:gan}. The training process based on the competition between the generator and the discriminator helps the generator to learn for generating increasingly realistic attack traffic flow. After a fine-tuning and training, our generator obtains the capability of crafting adversarial traffic flow that bypasses ML-based NIDS with high accuracy while maintaining the domain constraints. To enforce the domain constraints, we apply two types of enforcement: (1) Enforcement 1 ensures 
the inherent properties of the attack traffic is embedded in the generator training, (2) Enforcement 2 ensures 
the valid ranges of attack features is applied to the adversarial example after the perturbations applied to the traffic, as shown in Figure~\ref{fig:attack}.





\subsection{Constraints}
As previously discussed, NIDSGAN needs to generate adversarial traffic flows that comply with domain constraints, i.e., create realizable yet malicious network flows.
We consider traffic to be \textit{adversarial} if it is undetectable by the NIDS and we consider traffic to be \textit{constraint-compliant} if it is realizable in practice. The former is readily enforced by GANs and we enforce the latter through two types of constraints: (1) inherent constraints, and (2) valid ranges.

{}

The inherent constraints of network traffic flow are characterized by the traffic's protocol and attack type. Without constraints, the adversary could perturb $2^n$ (n-number of feature of traffic flow) different subsets of all the features. However, the network domain allows a very small number of subsets as valid perturbation for a given network traffic flow because the network traffic flows must be realizable in the real world. For instance, a network TCP flow (packet) may only certain flags and services, and UDP flows may have others. 

The constraint of valid ranges is critical for the adversary. Each feature in the adversarial traffic flow has its corresponding valid range. Out-of-range feature values render the attack invalid, as the traffic (1) loses its attack semantics, and (2) cannot be realized in practice. Therefore, extreme values out of the feature distributions can easily expose the attack traffic to NIDS. 

We explain the two constraints in detail along with their implementations in the following.

1) Inherent constraints
    \begin{enumerate}[label=(\alph*)]
        \item Attack type dictates that a set of feature values of traffic that must be kept constant. This preserves the underlying semantics of an attack.
        \item Protocol type dictates that a certain subset of features must be zero, as some features are protocol-specific (such as TCP flags). Also, a certain protocol is allowed to have only certain flags and services for feature values. 
        \item One-hot features converted from categorical must not be perturbed. This ensures perturbations applied by GANs result in realizable traffic.
    \end{enumerate}

\begin{table}
\centering
\caption{Depiction of Enforcement 1. \label{tab4}}
\scalebox{0.8}{
{\begin{tabular}{cccccccc} 
\hline
Vector Type/Index & 1 & 2 & 3 & 4 & 5 & ... &  n\\ 
\hline
Perturbation ($\Delta \vec{x}$)  & $\Delta \vec{x}_1$ & $\Delta \vec{x}_2$ & $\Delta \vec{x}_3$ & $\Delta \vec{x}_4$ & $\Delta \vec{x}_5$ &... & $\Delta \vec{x}_{n}$ \\
Mask Vector ($\vec{m}$)& 1 & 0 & 1 & 0 & 1 & ... & 1 \\
Result ($\Delta \vec{x} \circ \vec{m}$)& $\Delta \vec{x}_1$ & 0 & $\Delta \vec{x}_3$ & 0 & $\Delta \vec{x}_5$ &... & $\Delta \vec{x}_{n}$ \\
\hline
\end{tabular}}{}}
\end{table}

To enforce the first inherent constraint, we mask the computed perturbations on the corresponding features of all types ((a), (b), and (c)) at each iteration of the training as shown in Table~\ref{tab4}. Here, the perturbation computed by the generator is $\Delta \vec{x} = (\Delta x_1, ..., \Delta x_{n})^T$. (For the NSL-KDD and the CICIDS-2017 dataset, $n=123$ and $n=82$, respectively). We do this by taking the Hadamard product of the perturbation vector $\Delta \vec{x}$ with a mask vector $\vec{m}$ that enforces these constraints as described in the Equation~\ref{eqn:hadamard}. We implement this functionality within the function of the generator in the code.

 \begin{equation} 
    \begin{aligned}
    \label{eqn:hadamard}
    & \vec{m} = (m_1,...,m_{n})^T
    & m_k=\left\{
    \begin{array}{@{}ll@{}}
    1, \quad \textrm{feature included} \\
    0, \quad \textrm{otherwise}
    \end{array}\right.
    \end{aligned}
\end{equation}

    \begin{figure}[h]
    \centering
    \includegraphics[width=0.45\textwidth]{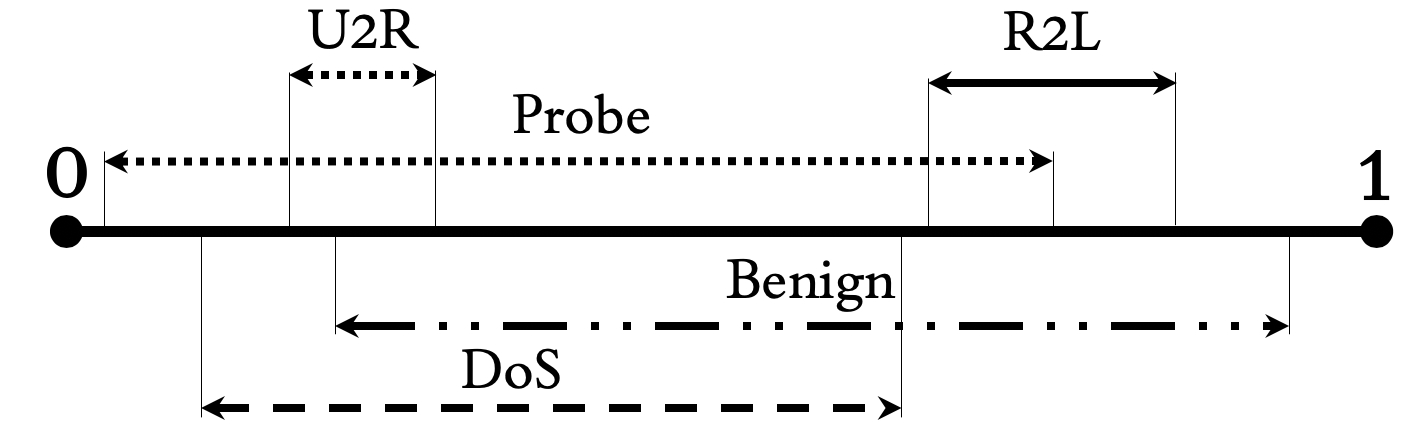}
    \caption{Enforcement 2: In the preprocessing, all the numerical features are scaled between 0 and 1. The valid ranges of \texttt{DoS}, \texttt{Probe}, \texttt{Benign}, \texttt{R2L}(Remote-to-Local), and \texttt{U2R} (User-to-Root) traffic flows are depicted.} 
    \label{fig:enforce2}
    \end{figure}
    
~

2) Valid ranges
        \begin{itemize}
        \item For each type of attack, we find the minimum and the maximum value of each feature and define the range between the minimum and maximum as the valid range. Then, after applying the perturbation to original attack traffic during the process of crafting adversarial examples, we enforce the valid range for each feature of attack traffic by clipping. Figure~\ref{fig:enforce2} shows that valid ranges of each class in NSL-KDD dataset for a particular feature $x_1$. For example, if the feature value of $x_1$ of a DoS attack traffic becomes $x_1 + \Delta x_1 < d_\textrm{min}$ after a perturbation is applied, then we clip the value and replace it with $d_\textrm{min}$.
    \end{itemize}

\subsection{Attack Flow Generation}

The main objective of our attack algorithm is to add a minimum possible perturbation to an attack traffic flow such that the resulting adversarial traffic flow will successfully bypass the DNN-based NIDS while maintaining its domain constraints. We accomplish this objective by training the optimal generator by optimizing the generator's loss function $L_\textrm{G}$ as shown in the following equation. 

\begin{equation} 
    \begin{aligned}
    & L_\textrm{G} = L_\textrm{adversarial} + \alpha L_\textrm{wgan-gp} + \beta L_\textrm{perturbation} \\
    \end{aligned}
    \label{eqn:gan}
\end{equation}

\noindent To achieve this, we leverage the AdvGAN~\cite{advgan} structure and make two main modifications to it in crafting adversarial traffic flows: 1) enforcing constraints: In each iteration of the algorithm training, we enforce both types of constraints as defined in last subsection. 2) improved GAN framework (Wasserstein GAN~\cite{wgangp} with gradient penalty (WGAN-GP) for the $L_\textrm{wgan-gp}$ term in Equation~\ref{eqn:gan}. WGAN-GP handles the problems of gradient explosion or vanishing, mode collapse, and non-convergence issues. Thus, it improves the performance of crafting adversarial traffic flows. For the original GANs' structure, two-loss functions are defined: $L_\textrm{G}$ for the generator and $L_\textrm{D}$ for the discriminator.

The detailed explanation of each term in the generator's loss function is in the following as described in Equation. ~\ref{eqn:gan}.

\begin{enumerate}
    \item The objective of $L_{\textrm{adversarial}}$ is to update the generator weights such that the perturbed attack traffic is classified as benign traffic flow. It measures how far $\vec{x}^* = \vec{x}+\Delta \vec{x}$ is from the boundary of the benign class. If the crafted adversarial traffic flow $\vec{x}^*$ is classified by the target NIDS as benign, the loss of this term will be 0. Otherwise, it penalizes the loss function based on cross-entropy. 
    
    \begin{equation} 
    \begin{aligned}
    & L_\textrm{adversarial} = \underset{\vec{x} \sim \mathbb{P}_r}{\mathbb{E}} [\textrm{Loss}(F(\vec{x}+\Delta \vec{x}),t)]
    \end{aligned}
    \end{equation}
    
    \item The objective of $L_\textrm{wgan-gp}$ is to make the perturbed attack traffic flow $\vec{x}^*$ to be indistinguishable from the original attack traffic flow. The discriminator D tries to determine the perturbed traffic flow $\vec{x}^*$ from the real traffic flow. This helps in two ways: 1) it helps to minimize the perturbation 2) it helps to maintain the domain constraints. Two types of GAN losses are used in the experiments. 
    The $L_\textrm{wgan-gp}$ represents the Wasserstein GAN with gradient penalty loss \cite{wgangp}. Therefore, WGAN-GP loss provides better results in our experiments. In the following Equation~\ref{eqn:wgan-gp}, the first and second terms correspond to the losses of the discriminator and the generator, respectively, while the third term represents the gradient penalty term (the details of the term is shown in Algorithm 1).
    
    
     \begin{equation} 
    \begin{aligned}
     L_\textrm{wgan-gp} = \min_G \max_D  \left\{ \underset{\vec{x} \sim \mathbb{P}_r}{\mathbb{E}}   [D(\vec{x})] - \underset{\vec{x} \sim \mathbb{P}_r}{\mathbb{E}}  [D(\vec{x}^*)] \right. \\
     \left. + \underset{\hat{x}}{\mathbb{E}} [\lambda (\| \nabla_{\hat{x}} D(\hat{x})\|_2-1)^2] \right\} 
        \label{eqn:wgan-gp}
    \end{aligned}
    \end{equation}

    \item  The objective of $L_\textrm{perturbation}$ is to minimize the perturbation $\Delta \vec{x}$ so that the perturbed traffic will be as close as the real input traffic flow. With $L_\textrm{wgan-gp}$ optimization, the distribution of $\vec{x}+\Delta \vec{x}$ approaches that of the real traffic flow~\cite{gan1}. As specified in most of the prior work in the AML community, we bound the $l^2$ perturbation magnitude of the loss function with the following equation where the $\epsilon-$bound is typically chosen to be 0.3. Hinge loss is commonly used for the optimization of the perturbation. Therefore, optimizing for the hinge loss results a better success rate than optimizing for $L_\textrm{perturbation} = \| \Delta \vec{x} \|_2$.
    
    \begin{equation} 
    \begin{aligned}
    & L_\textrm{perturbation} = \underset{\vec{x} \sim \mathbb{P}_r}{\mathbb{E}} [ \max(0, \| \Delta \vec{x} \|_2 - \epsilon)]
    \end{aligned}
    \end{equation}
\end{enumerate}


\begin{table}[htbp]
\centering
\caption{Parameters for the NIDSGAN training for generating the adversarial network traffic flows. \label{tab3}}
\scalebox{0.75}{
{\begin{tabular}{cccccc} 
\hline

Dataset & NIDSGAN & Batch & Adam & Epochs & Time\\
& Coefficient & Size & Optimizer \\
\hline
CICIDS & $\alpha=0.1,\beta=0.2$ & $20x$ & $\alpha=10^{-3},\beta_1=0.5,\beta_2=0.9$  & $800$ & $~1.5 h$ \\
NSL-KDD & $\alpha=0.1,\beta=0.2$ & $28x$ & $\alpha=10^{-3},\beta_1=0.5,\beta_2=0.9$ & $800$ & $~1.2 h$ \\
\hline
\end{tabular}}{}}
\end{table}

The optimal values of $\alpha$ and $\beta$ coefficients in the generator's loss function are found to be 0.1 and 0.2 with a systematic grid search method. Algorithm~\ref{alg:NIDSGAN} shows the algorithmic details for crafting adversarial examples that bypass DNN-based NIDS.

\begin{algorithm}
\caption{NIDSGAN: Crafting adversarial network traffic flows that bypass DNN-based NIDS}
\begin{algorithmic}
\REQUIRE  {$x$-input, $F$-target NIDS, $G$-generator, $D$-discriminator, $t$-target class, $E_1$-enforcing type-1 constraints, $E_2$-enforcing type-2 constraints, $n$-number of epochs, $m$-number of batches}
\REQUIRE  {initial discriminator parameters $w_0$, initial generator parameters $\theta_0$}

\WHILE{$\arg \max F(\vec{x}^*) \neq t$}{
    \FOR{epoch=1,...,n}{
        \FOR{i=1,...,m}{
            \STATE Sample real data $\vec{x} \sim  \mathbb{P}_r$.
            \STATE  $ \quad \quad \Delta \vec{x} \leftarrow \textrm{Enforce}_1(G(\vec{x}))$\;
            \STATE   $ \quad \quad  \vec{x}^* \leftarrow \textrm{Enforce}_2(\vec{x} + \Delta \vec{x}$)\; 
            \STATE   $ \quad  \quad \hat{x} \leftarrow \sigma \vec{x} + (1-\sigma) \vec{x}^*$\;
            \STATE $L_\textrm{wgan-gp} \leftarrow D(\vec{x})-D(\vec{x}^*)+\lambda (\| \nabla_{\hat{x}} D(\hat{x})\|_2-1)^2$\;
            \STATE $ w \leftarrow \textrm{AdamOptimizer}(\nabla_w  L_\textrm{wgan-gp}^{(i)}, w, \alpha)$\;
            \STATE $ \theta \leftarrow \textrm{AdamOptimizer}(\nabla_\theta  L_\textrm{wgan-gp}^{(i)} + L_\textrm{Pert}^{(i)} + L_\textrm{Adv}^{(i)}, \theta, \alpha)$
            }\ENDFOR
        }\ENDFOR
}\ENDWHILE
\label{alg:NIDSGAN}
\end{algorithmic}
\end{algorithm}

\subsection{Active Learning}

Active learning ~\cite{activeLearning} is a ML training technique extensively used for scenarios where no sufficient amount of labeled training data is available. In the adversarial machine learning field, using the intuition of active learning, the attacker can approximate a decision boundary of the target model by intelligently probing it. However, this technique has never before been used for constrained domains. In this context, we use an active learning technique to amplify our success rates on a restricted blackbox model with a limited size of training data associated with the adversary's local model. We can achieve this by iteratively evaluating the classification of samples that are near the decision boundary of the target model. Applying the active learning technique for crafting attack traffic flows is more challenging than applying it to the image domain because the constraints of the network domain significantly limit the allowed perturbation space.

\begin{figure*}[h]
    \centering
    \includegraphics[width=0.7\textwidth]{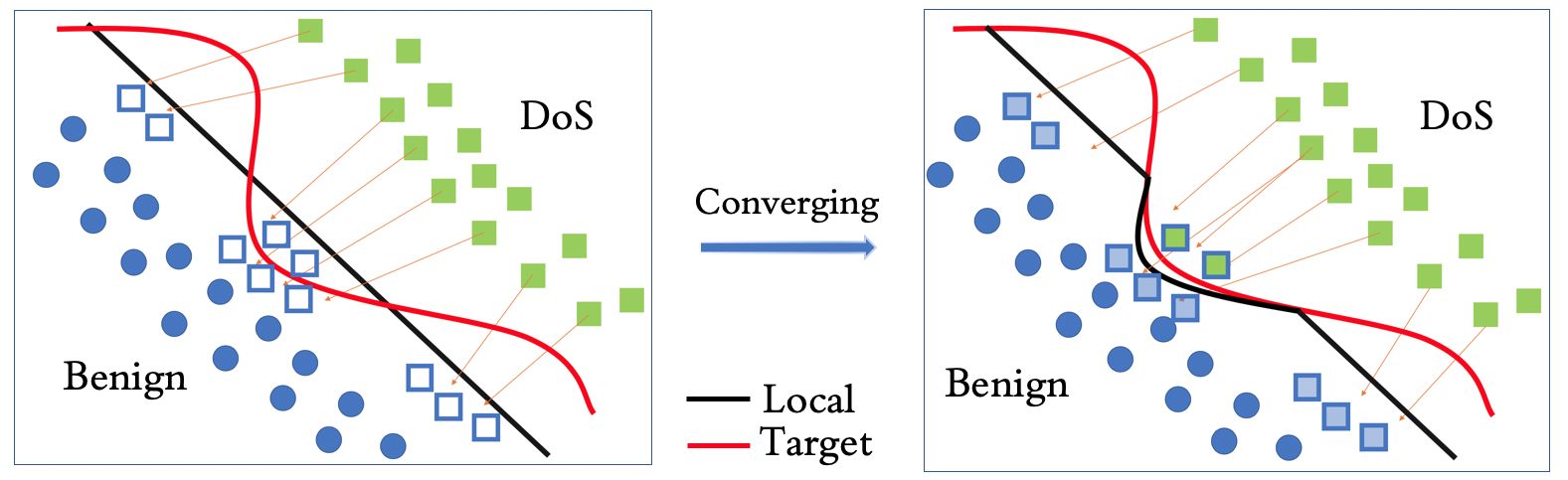}
    \caption{Learning decision boundary of a target model through active learning}
    \label{fig:ActiveLearning}
\end{figure*}

We illustratively depict how active learning helps with aligning the decision boundaries between the target and local models in Figure~\ref{fig:ActiveLearning}. Red and black lines in the figure represent the imaginary decision boundaries (between \texttt{DoS} and the \texttt{Benign}) of the target and the local model, respectively. In this example, two out of ten crafted adversarial examples that successfully evade the local model fail to bypass the target model (i.e., two empty squares on the right side of the red boundary in Figure~\ref{fig:ActiveLearning}) by the target model although all of them bypass the local model. To align decision boundaries, we use these two particular samples with their labels from the target model to finely adjust the regional decision boundary of the local model. As a result, the regional decision boundary of the local model is similarized to the target model, so that the two previously failed adversarial samples can now successfully evade the target model.

The main difference in our process compared with most of the previous work in blackbox setting\cite{advgan} is that we use labels classified by the target model instead of the probability vector provided by the target model. Using only labels is more realistic and challenging to craft adversarial examples. Knowing the probability vector of the target model gives much more information about the decision boundary of the target model; therefore, it is much easier to learn the decision boundary of the target model for the adversary.


\subsection{Realizing Adversarial Flows}

The last phase of our attack is to execute PCAPs over the network such that they bypass NIDS with real traffic, i.e., realize the malicious intent of the adversarial example.  
There are several technical challenges in creating ``playback'' for an adversarial flow , including packet timing, traffic shaping, and other techniques. 
As it turns out there are several well-known tools that serve this purpose, such as tcpreplay~\cite{Tcpreplay}, Bit-Twist~\cite{Bit-Twist}, and Packet Player~\cite{Colasoft}. They could be used to replay edited PCAP data to generate the adversarial network traffic with expected high fidelity. Moreover, some works have demonstrated success in exclusively perturbing statistical information (i.e., perturbations to features that have no constraints) so that network packets can be replayed to bypass NIDS~\cite{homoliak2019improving,hashemi2019towards}. Either of these frameworks, with some necessary extensions beyond perturbations on statistical features, could be leveraged by our approach to produce a network flow that is representative of the generated adversarial examples.  We are exploring executing these tools on the real networks (which is a complex and lengthy exercise), but for brevity we defer those experiments to our future work. We defer interested readers to Table~\ref{tab:mapback} in the Appendix on how to apply each of the cited techniques to realize our adversarial flows.

%% file: 5.Evaluation.tex
\section{Evaluation}
In this section, we evaluate NIDSGAN against two DNN-based NIDS in the literature as well as an exemplar NIDS model trained on NSL-KDD and CICIDS-2017 datasets. We consider traditional whitebox and blackbox threat models in the literature as well as a more realistic restricted-blackbox threat model in our experiments. We also discuss that the success rate of our algorithm can be significantly amplified by an active learning technique and show that it is most effective and important in the restricted-blackbox threat models for the adversary. We also look at classical NIDS trained with classical ML algorithms and show that they are vulnerable to adversarial traffic flows generated by the NIDSGAN. 

\subsection{Datasets}
\textbf{NSL-KDD} ~\cite{nsl-kdd1} dataset is arguably the most popular dataset in the intrusion detection community. It is an improved version of KDD Cup99, which includes redundant records in both training and testing sets. NSL-KDD dataset has five classes: \texttt{DoS}, \texttt{Probe}, \texttt{R2L}, \texttt{U2R}, and \texttt{Benign}. The dataset contains 125,973 and 22,543 samples for training and testing, respectively. Each sample has 41 features (123 after one-hot encoding) that can be divided into four different categories: basic, content, time-based, and host-based features. (1-9) basic features contain the intrinsic information about the packet such as `protocol type', `duration', `service', and `flag'; (10-22) content features contain original packets' information such as `number of failed logins' and `root shell'; (23-31) time-based features are traffic input computed by two-seconds time window including `count', `srv count', and `same srv rate'; (32-41) host-based features are traffic features for analyzing attacks from destination to host looking at time-windows that last longer than two-second such as `dst host count' and `dst host srv serror rate'.

\textbf{CICIDS-2017}~\cite{cicids-2017} dataset consists of network traffic flows simulated within five weekdays. The dataset is simulation-based from 25 internet users including their email, HTTP, HTTPS, FTP and SSH traffic stored in PCAP format. The dataset consists of 14 classes. We merge the 14 classes into 7 finer-grained classes by combining similar types of attacks as recommended in ~\cite{cicids2} to balance the distribution of the classes and improve the classification accuracy. These 7 classes are \texttt{Benign}, \texttt{Bot}, \texttt{Pat} (Patator), \texttt{DoS}, \texttt{Inf} (Infiltration), \texttt{Port} {port-scan}, and \texttt{Web}. The dataset is divided into training and test sets with a ratio of 75\% and 25\%, respectively. There are 82 features in the dataset after preprocessing.

\subsection{Target NIDS for Detecting Attack Flows}
In this work, we evaluate our attack algorithm against two DNN-based NIDS models in the literature as well as our (self-created) exemplar NIDS model. We choose the AlertNet~\cite{alertnet} and the DeepNet~\cite{deepnet} as shown in Tables~\ref{model1} and \ref{model2} because these models provide the highest accuracy as well as the lowest false positives and false negatives compared to other existing models. The layers of the NIDS' DNN models consist of fully connected perceptrons for the main structures, employing ReLU activation functions which provide non-linearity, batch normalizations for improving training performance, and dropouts for regularization to avoid overfitting.

\newcommand\ppp{\phantom{00000}}

\begin{table}[htbp]
\centering
\caption{AlertNet ~\cite{alertnet} DNN model \label{model1}}
\scalebox{0.75}{
{\begin{tabular}{ccccccc} 
\hline
Type/Layer & First & Second & Third & Fourth & Fifth & Output\\ 
\hline
MLP (ReLU)           & 1024 & 768 & 512 & 256 & 128 & - \\
Batch Normalization  & 1024 & 768 & 512 & 256 & 128 & - \\
Dropout (0.01)       & 1024 & 768 & 512 & 256 & 128 & - \\
Fully Connected (Softmax)     & - & - & - & - & - & 7 or 5 \\

\hline
\end{tabular}}{}}
\end{table}

\begin{table}[htbp]
\centering
\caption{DeepNet ~\cite{deepnet}  DNN model \label{model2}}
\scalebox{0.75}{
{\begin{tabular}{ccccccc} 
\hline
Type/Layer & First & Second & Third & Fourth & Output\\ 
\hline
MLP (ReLU)           & 256 & 256 & 256 & 256 & - \\
Dropout (0.01)       & 256 & 256 & 256 & 256 & - \\
Fully Connected (Softmax)     & - & - & - & - & 7 or 5 \\

\hline
\end{tabular}}{}}
\end{table}

To train our exemplar DNN model (IdsNet), we vary the number of layers and the number of neurons in the layers as shown in Table~\ref{tab1} and \ref{tab2}. Theoretically, the optimal number of hidden layers of the DNN is related to the dataset and task complexity ~\cite{pedrycz2020development}. In this work, we choose an optimal architecture of a target NIDS for each dataset based on the evaluations as shown in Table. ~\ref{tab1} and \ref{tab2}. We experimented with five different DNNs differed by their number of layers. Each layer of DNNs consists of several perceptrons, a ReLU activation function, and a dropout function that prevents overfitting. For each case, we calculate their accuracy, precision, recall, and F1-score as shown in Table.~\ref{tab1} and \ref{tab2} (where one-layer DNN represents a multiclass logistic regression).

For choosing the number of perceptrons in the hidden layers, we used the benchmark rule that the number of perceptrons in the hidden layers should be between the number of neurons in the input and output layer (the number of perceptrons is modified for performance improvement) ~\cite{heaton2008introduction}. After several iterations, we find that the following architectures can provide results as high as the results shown in the NIDS 
proposed by the previous work~\cite{best_nsl_1, best_nsl_2, cicids2, deepnet}. As shown from the experiments in Table. ~\ref{tab1} and ~\ref{tab2}, adding more layers to 3-layer DNN does not improve the accuracy due to overfitting, but it increases the computational cost. The same conclusion was made in the previous work~\cite{alertnet}.

\begin{table}[htbp]
\centering
\caption{IdsNet for NSL-KDD. \label{tab1}}
\scalebox{0.75}{
{\begin{tabular}{ccccccc} 
\hline
Architecture & Layers & Accuracy & Precision & Recall & F-score\\ 
\hline
DNN 1 layer   & 123,5 &   0.75 & 0.89 & 0.62 & 0.74 \\
DNN 2 layers  & 123,64,5 &   0.78 & 0.89 & 0.65 & 0.75  \\
DNN 3 layers  & 123,64,32,5 &   0.79 & 0.89 & 0.65 & 0.75   \\
DNN 4 layers  & 123,80,64,24,5 &   0.77 & 0.88 & 0.63 & 0.73  \\
DNN 5 layers  & 123,80,48,32,16,5 &   0.77 & 0.88 & 0.63 & 0.74  \\
\hline
\end{tabular}}{}}
\end{table}

\begin{table}[htbp]
\centering
\caption{IdsNet for CICIDS-2017. \label{tab2}}
\scalebox{0.75}{
{\begin{tabular}{ccccccc} 
\hline
Architecture & Layers & Accuracy & Precision & Recall & F-score\\ 
\hline
DNN 1 layer   & 82,7 & 0.97 & 0.88 & 0.97  & 0.92 \\
DNN 2 layers  & 82,42,7 &  0.980 & 0.928 & 0.964 & 0.946 \\
DNN 3 layers  & 82,42,21,7 & 0.985  & 0.949 &  0.969 &  0.959\\
DNN 4 layers  & 82,60,42,28,7 & 0.982  & 0.942  & 0.958  & 0.950 \\
DNN 5 layers  & 82,60,42,28,14,7 & 0.96 &0.91& 0.87 &0.89 \\
\hline
\end{tabular}}{}}
\end{table}

For these reasons, we selected the three-layer DNN as our target for the NIDS experiments. The prediction accuracy, architecture, and hyperparameters of the target models' are shown in Table. ~\ref{tab3}. According to published results of the DNN-models trained on CICIDS and NSL-KDD datasets, the prediction accuracies we found in our models are at the same level as the state-of-the-art predictions~\cite{alertnet}. In the next section, we introduce domain constraints. 

\begin{table}[htbp]
\centering
\caption{Prediction Set Up for Target Models of Our Exemplar (IdsNet). \label{tab3}}
\scalebox{0.75}{
{\begin{tabular}{cccccc} 
\hline
Dataset & Layers (MLP) & Batch Size & Learning Rate & Epochs & Accuracy\\ 
\hline
CICIDS & 82,42,21,7 & 256 & 0.001 & 50 & 98.5\% \\
NSL-KDD & 123,64,32,5 & 32 & 0.01 & 50 & 79.8\% \\
\hline
\end{tabular}}{}}
\end{table}

\subsection{Whitebox Threat Model}
\begin{figure}[h]
    \centering
    \includegraphics[width=0.49\textwidth]{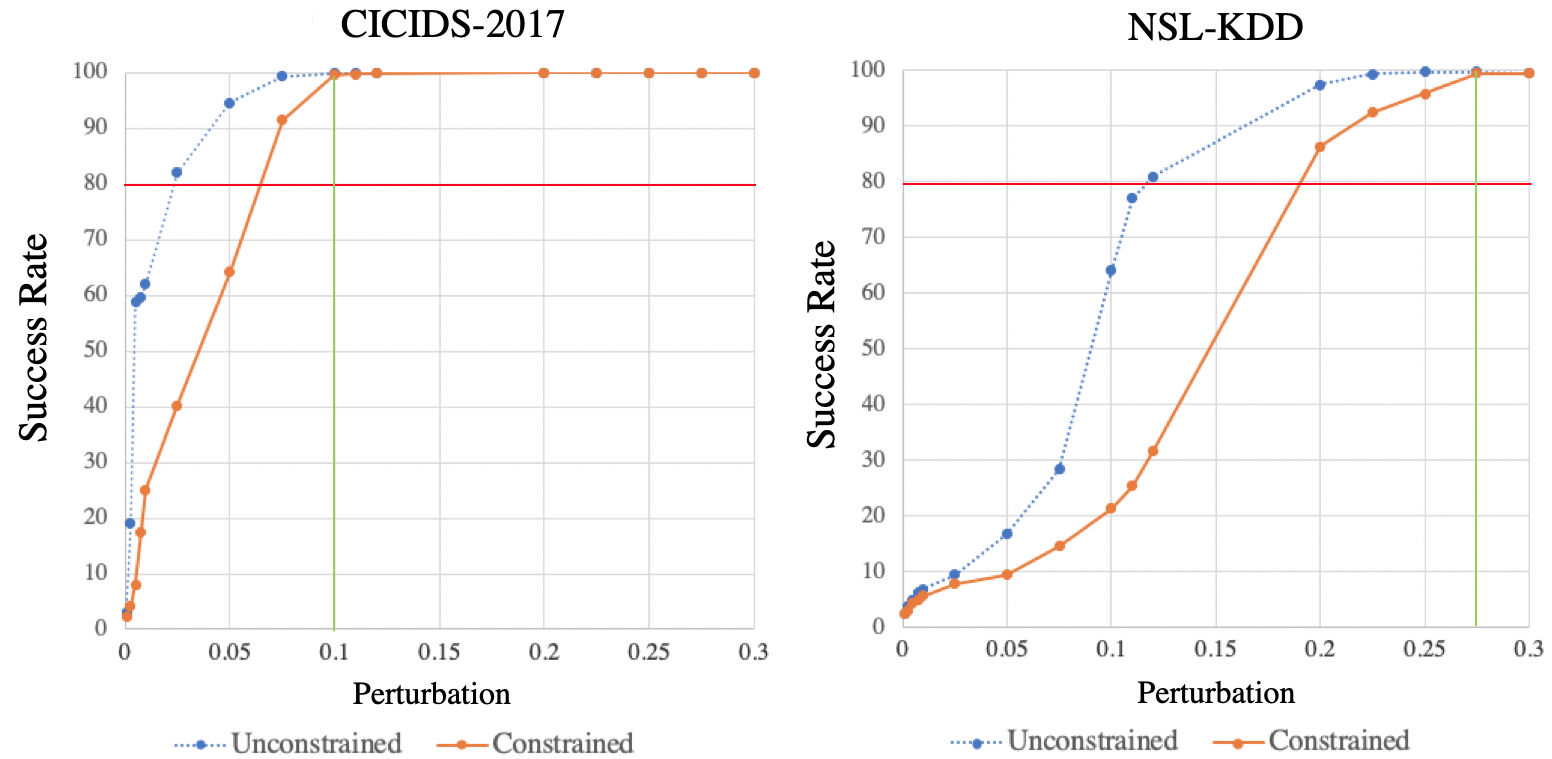}
    \caption{Success rate comparison of adversarial network traffic flows generated on CICIDS-2017 and NSL-KDD dataset in unconstrained and constrained cases.}
    \label{fig:final}
\end{figure}

Whitebox threat model in these experiments represents that the adversary has full access to the training data of the NIDS models. In the first set of experiments, we seek to analyze the effect of the domain constraints on our attack algorithm. Here we generate adversarial samples using the NIDSGAN and vary the amount of allowable perturbation, as shown in Figure~\ref{fig:final}.  We find that generating adversarial traffic flows is more challenging with constraints enforced. The red line in Figure~\ref{fig:final} shows that the constrained case requires two times larger perturbation than the unconstrained case to achieve the same 80\% success rate in both CICIDS-2017 and NSL-KDD cases. In other words, restricting the perturbation applied to important features for the classification costs larger perturbations applied to the less important features for the classification. Furthermore, we can see that the effect of the domain constraints is large when the perturbation is small. 

%

It is more challenging to bypass NIDS models trained with NSL-KDD. The vertical green lines show that to achieve a nearly 100\% success rate, the CICIDS-2017 case requires 0.1 perturbation while the NSL-KDD case requires a much larger 0.275 perturbation. The number of classes (7 vs 5), percentage benign traffic flows in the dataset (83\% vs 52\%), and more importantly, the flexibility of constraints directly dependent on dataset sizes (3 million vs 125 thousand samples) are the main factors that result in this phenomenon, which is to be explained in more detail in the blackbox section. 

More broadly NIDSGAN  is an effective method generating adversarial traffic flows in the whitebox threat model achieving nearly 100\% success rates with less than 0.275 perturbations in both CICIDS-2017 and NSL-KDD cases. The changes done to the traffic flow features is significantly small. For instance, $\|\Delta \vec{x}\|_2$=0.275 can be a perturbation applied to a single feature out of a total of 82 (123) features of the traffic flow. If there are two features perturbed, the changes to the features can be 0.21. If there are 10 features perturbed, the average change to the feature is only 0.1, etc. A more detailed lookup table is shown in Appendix Table~\ref{tab:pertMeasure}. 

\begin{figure}[h]
    \centering
    \includegraphics[width=0.49\textwidth]{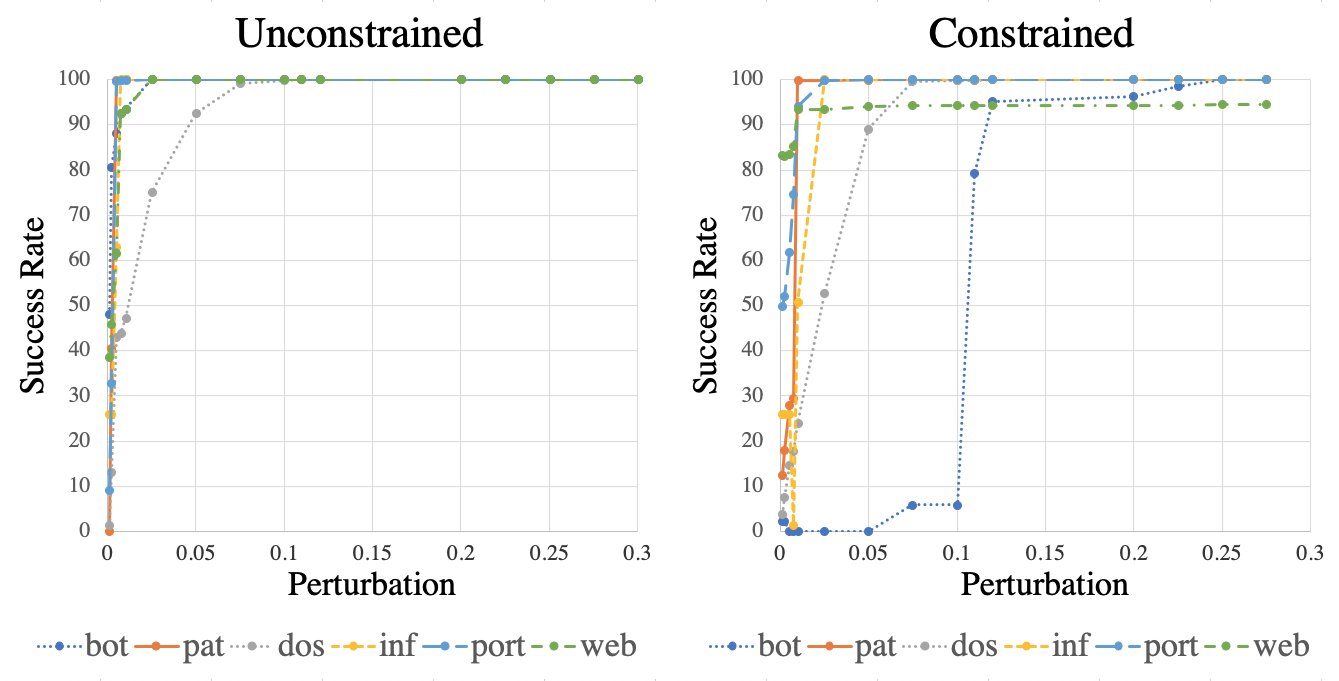}
    \caption{Comparison of success rates for each attack evaluated against IdsNet trained with CICIDS-2017 dataset in unconstrained and constrained cases.}
    \label{fig:ids_com}
\end{figure}

\begin{figure}[h]
    \centering
    \includegraphics[width=0.49\textwidth]{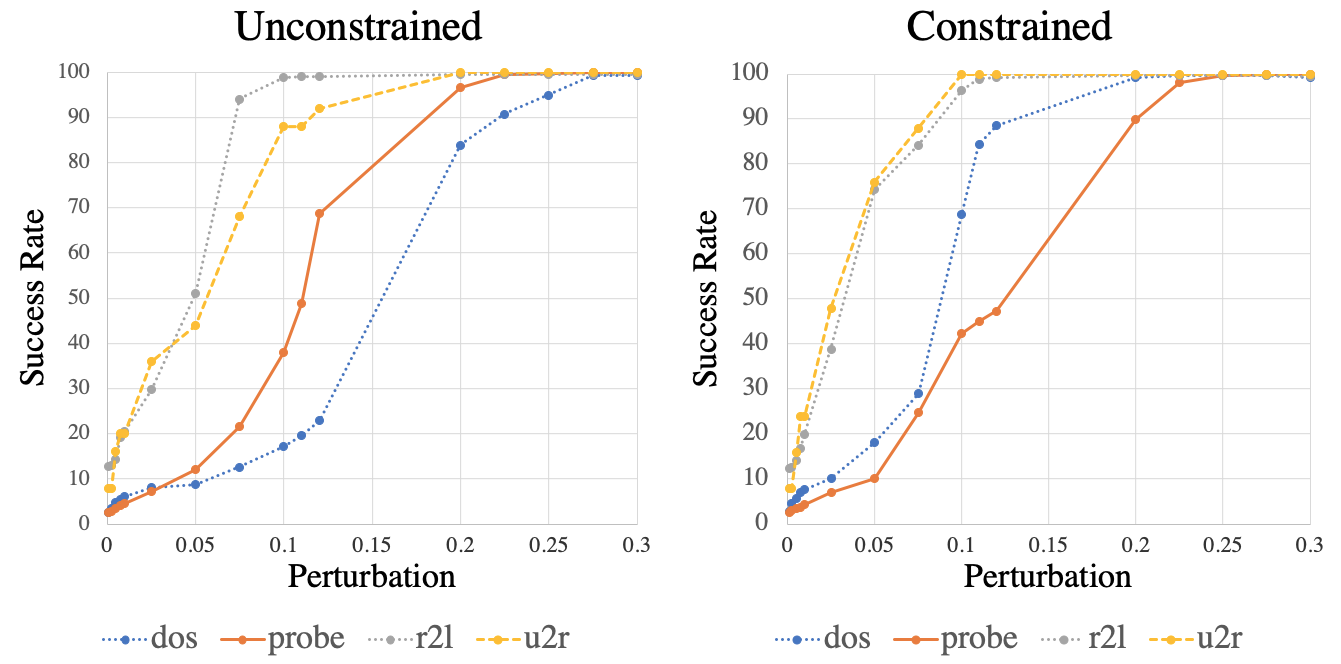}
    \caption{Comparison of success rates for each attack evaluated against IdsNet trained with NSL-KDD dataset in unconstrained and constrained cases.}
    \label{fig:nsl_com}
\end{figure}

To show the effect of adding constraints for each attack class, success rate comparisons of different attack categories between unconstrained and constrained cases are plotted in Figure~\ref{fig:ids_com} and ~\ref{fig:nsl_com}. Our algorithm achieves nearly 100\% success rate with perturbation size less than 0.1 for \texttt{pat}, \texttt{dos}, \texttt{inf}, and \texttt{port} attacks. For web attacks, the highest success rate we can achieve is 95\%. For \texttt{bot} attacks, the success rate reaches only 5\% when $\|\Delta \vec{x}\|_2$ perturbation is less than 0.1. The main reason for this phenomenon is that the \texttt{bot} and the \texttt{web} attack traffic are the second and third smallest attacks in the CICIDS dataset behind \texttt{inf} attack traffic in terms of class size. Therefore, the constraint enforced by our algorithm eliminates the huge chunk of perturbation space that can be applied in an unconstrained domain.

With the NSL-KDD dataset, generating adversarial traffic flows gets more challenging with constraints. Compared to CICIDS-2017, at least a perturbation level of 0.2 is needed to achieve success rates of more than 90\% for all attacks. Enforcing the constraints of the attacker is more challenging for the NSL-KDD dataset. One of the main reasons is that CICIDS-2017 has more data samples; therefore, it provides a larger perturbation space. 

\textbf{Takeaway:}
 The results in the whitebox setting demonstrate that our algorithm is effective in crafting adversarial traffic flows, achieving near 100\% success rates with $l^2$ norm of only $\| \Delta \vec{x}\|_2  = 0.275$ as shown in Figure~\ref{fig:final}. They also show the challenge of taking domain constraints into account by slower convergence and larger required size of perturbations. However, the assumptions of the whitebox threat model are ideal for the adversary because it assumes that the adversary has access to the full datasets or the model internals of the NIDS models. For that reason, we consider more constrained realistic threat models in the section. 

\subsection{Blackbox Threat Model}

Next, we evaluate NIDSGAN in a blackbox setting.   We begin by randomly sampling 10k network traffic flows ($<$7\% of NSL-KDD training data and $<$1\% of CICIDS-2017 training data) from the original training set (and keeping the class balance fixed). We then evaluate the adversarial success rates\footnote{The adversarial success rate is the percentage of perturbed flows whose label flips from malicious to benign under the NIDS model.} based on subsets of these 10k records as the adversary's local model training.  Note that we consider much more limited access to model training data in the following section.

\begin{table}
\centering
\caption{GAN vs WGAN-GP in blackbox setting $\| \Delta x \|_2$ less than 0.3\label{tab:ganVSwgan}}
\scalebox{0.9}
{\begin{tabular}{ccccc} 
\hline
Local Model's Set & \multicolumn{2}{c}{\textbf{NSL-KDD}} & \multicolumn{2}{c}{\textbf{CICIDS-2017}}\\ 
\cline{2-5} 
out of 10k & GAN & \textbf{WGAN-GP} & GAN & \textbf{WGAN-GP}\\ 
\hline
1.4\% & 0.14 & \textbf{0.19} & 0.88 & \textbf{0.90}\\ 
2.8\% & 0.21 & \textbf{0.39} & 0.91 & \textbf{0.92} \\
5.6\% & 0.44 & \textbf{0.44} & 0.92 & \textbf{0.97}  \\
7.4\% & 0.52 & \textbf{0.60} & 0.95 & \textbf{0.98} \\
12.0\% & 0.52 & \textbf{0.75} & 0.98 & \textbf{0.99} \\ 
24.0\% & 0.61 & \textbf{0.80} & 0.99 & \textbf{0.99} \\
36.0\% & 0.66 & \textbf{0.81} & 0.99 & \textbf{0.99}  \\
67.0\% & 0.87 & \textbf{0.89} & 0.99 & \textbf{0.99} \\
100.0\% & 0.91 & \textbf{0.96} & 0.99 & \textbf{0.99} \\ 
\hline
\multicolumn{5}{c}{IdsNet}\\
\end{tabular}}{}
\vspace{1ex}

     {\raggedright Adversarial success rate as a function of local training data set size.}

\end{table}

In Table.~\ref{tab:ganVSwgan}, we compare GAN-based and WGAN-GP-based attack algorithms as described in Equation~\ref{eqn:gan} and show that WGAN-GP consistently outperforms the original GAN on both datasets. Along with improving the gradient and non-convergence issues of GAN, we observe that WGAN-GP requires almost no hyperparameter tuning compared to GAN, which reduces the difficulty of training, especially with a small training size. Applying WGAN-GP in our algorithm gives the advantages over the GAN-based attack algorithms proposed in the previous works. 
\begin{table}
\centering
\caption{Adversarial success rate of the NIDSGAN in blackbox setting with $\| \Delta x \|_2$ less than 0.3 on NSL-KDD dataset. \label{blackbox_nsl}}
\scalebox{0.9}
{\begin{tabular}{cccc} 
\hline
Local Model's Set  & \multicolumn{3}{c}{\textbf{Success Rate}}\\ 
\cline{2-4} 
out of 10k & AlertNet & DeepNet & IdsNet  \\ 
\hline
1.4\%  & 0.27    & 0.27     & 0.19 \\ 
2.8\%  & 0.36   & 0.30    & 0.39  \\
5.6\%  & 0.41    & 0.33    & 0.44   \\
7.4\%  & 0.51    & 0.52     & 0.60 \\
12.0\% & 0.66    & 0.81     &0.75\\ 
24.0\% & 0.85    & 0.82     &0.80  \\
36.0\% & 0.88    & 0.83     &0.81   \\
67.0\% & 0.88    & 0.83     &0.89  \\
100.0\% & 0.89    & 0.85     &0.96  \\ 
\hline
\multicolumn{4}{c}{NSL-KDD}\\
\end{tabular}}{}
\vspace{1ex}

     {\raggedright The numbers in the first column represent the size of initial training with respect to the adversary's training set (10k) as well as the number of queries from the NIDS model. \par}

\end{table}

\begin{table}
\centering
\caption{Success rate of the NIDSGAN in blackbox setting with $\| \Delta x \|_2$ less than 0.3 on CICIDS-2017 dataset. \label{blackbox_ids}}
\scalebox{0.9}
{\begin{tabular}{rccc} 

\hline
Local Model's Set & \multicolumn{3}{c}{\textbf{Success Rate}}\\ 
\cline{2-4} 
out of 10k \phantom{00} & AlertNet & DeepNet & IdsNet  \\ 
\hline
1.4\% \ppp & 0.92 & 0.96 & 0.90 \\ 
2.8\% \ppp & 0.97 & 0.97 & 0.92  \\
5.6\% \ppp& 0.98 & 0.97 & 0.97   \\
7.4\% \ppp& 0.99 & 0.99 & 0.98 \\
12.0\% \ppp& 0.99 & 0.99 & 0.99 \\ 
24.0\% \ppp& 0.99 & 0.99 & 0.99  \\
36.0\% \ppp& 0.99 & 0.99 & 0.99   \\
67.0\% \ppp& 0.99 & 0.99 & 0.99  \\
100.0\% \ppp& 0.99 & 0.99 & 0.99  \\ 
\hline
\multicolumn{4}{c}{CICIDS-2017}\\
\end{tabular}}{}
\vspace{1ex}

     {\raggedright Adversarial success rate as a function of local training data set size.}

\end{table}

We evaluate our attack algorithm against two DNN-based NIDS models in prior work, as well as our well-trained model as shown in Table. ~\ref{blackbox_nsl} and ~\ref{blackbox_ids}. It is shown that with the entire local training set (10k/100\%) queries, the adversary can achieve success rates greater than 90\% on average for both datasets. To make the threat model even more restricted and challenging, we further consider cases where the adversary can launch as low as 140 traffic flows (1.4\% of the local training set) and show the success rates in Table. ~\ref{blackbox_nsl} and ~\ref{blackbox_ids}. In real-world scenarios, it is common to assume the adversary can only launch such a small number of traffic flows in the networks for two main reasons. 1) The adversary could easily be exposed by the NIDS for launching a large number of network attack traffic flows. 2) Finding a large number of traffic flows can be challenging for the adversary. The goal of an adversary in the practical blackbox threat model is to achieve a high attack success rate with a minimum number of network traffic flows for training the local model.

With only 1.4\% of the adversary's training set (or 140 traffic flows), the success rates for NSL-KDD and CICIDS-2017 datasets are more than 19\% and 90\% if WGAN-GP is used, respectively, as shown in Table. ~\ref{blackbox_ids} and ~\ref{blackbox_nsl}. With all of the adversary's training set (10k samples) for the local model's training, we can approximate the decision boundary of the target model well so that the success rate becomes as high as the success rate of the whitebox case.

\begin{table}[h]
\centering
\caption{Improvement with active learning technique on NSL-KDD dataset in the restricted-blackbox threat model. \label{activeLearning}}
\scalebox{0.65}
{\begin{tabular}{c|cccccc} 
\hline
Local Model's Set & \multicolumn{2}{c}{\textbf{AlertNet}} & \multicolumn{2}{c}{\textbf{DeepNet}}  & \multicolumn{2}{c}{\textbf{IdsNet}} \\ 
\cline{2-7} 
Out of 10k & Original & \textbf{A.Learning} & Original & \textbf{A.Learning} & Original & \textbf{A.Learning} \\ 
\hline
1.4\% & 0.27 & \textbf{0.37} & 0.27 & \textbf{0.32} & 0.19 & \textbf{0.25}  \\ 
2.8\% & 0.36 & \textbf{0.47} & 0.30 & \textbf{0.37} & 0.39 & \textbf{0.51}  \\
5.6\% & 0.41 & \textbf{0.63} & 0.33 & \textbf{0.47} & 0.44 & \textbf{0.57}  \\
7.4\% & 0.51 & \textbf{0.70} & 0.52 & \textbf{0.64} & 0.60 & \textbf{0.68}  \\
\hline
\multicolumn{7}{c}{NSL-KDD}\\
\end{tabular}}{}
     {\raggedright A.Learning column represents the increase in the success rate that comes with the cost of querying the adversarial examples crafted on the initial training size two times. \par}
\end{table}

Our blackbox attack is shown to be very effective on the CICIDS-2017 compared to the NSL-KDD. These two important factors favor the CICIDS-2017 dataset over the NSL-KDD dataset for the adversary: 1) CICIDS-2017 dataset has more number of classes as well as a higher percentage of benign network traffic flow than NSL-KDD dataset. Due to a much higher imbalanced dataset, it is easier to convert the classification of attack traffic flow to benign. (7 classes vs 5 classes and 80\% vs 53\% of benign traffic). 2) More importantly, the perturbation space is found to be much larger on the CICIDS-2017 dataset. These factors directly shape the complexity of the models' decision boundary and the allowed perturbation space by the constraints. 

To summarize the blackbox setting, we find that (1) our approach can achieve a success rate of over 80\% and 99\% on KDD and CICIDS datasets, respectively, with only less than 3600 (36\%) local model's training samples (2) the improved WGAN-GP framework outperforms the original GAN; and (3) CICIDS is consistently easier to evade due to larger perturbation space.




\subsection{Restricted Blackbox Attack}

Turning to the most realistic setting, we define the restricted blackbox model to be when the local model's training to less than or equal to 7.4\% of the adversary's blackbox set (or 740\footnote{The number 740 was selected because it represented a sample size that maintained class balance and a training dataset level where the existing blackbox attacker essentially ineffective (see the previous section).} samples) as shown in Table~\ref{blackbox_nsl} and ~\ref{blackbox_ids}. The results of the CICIDS-2017 case achieve more than 90\% success rate against all of the NIDS DNN models; however, the results of the NSL-KDD case are clearly not satisfactory. It is challenging for an adversary to find a sufficient number of network traffic flows to train the local model. Without a sufficient number of samples, the local model can not learn the decision boundary of the target model; therefore, the success rate of the attack is poor. In this situation, the adversary can improve the success rate by iteratively augmenting the initial scarce training set with active learning as shown in Table~\ref{activeLearning}. Results show that active learning significantly improves the success rate, by as high as 22\%, 14\%, and 13\% evaluated against AlertNet, DeepNet, and IdsNet, respectively when the number of the local models training set is 560 (or 5.6\%). 

The active learning works much more effective with a small local model's training size ($\leq$ 7.4\% of the adversary's set) than with a large local model's training size for two reasons: (1) local model trained with a small number of samples cannot approximate the decision boundary of the target model well; active learning helps to tune the local models' decision boundary with additional information from the target model. (2) as the local model's training samples increases, the use of additional information gained from the target model decreases because the existing samples' territory overlaps the adversarial samples created during the active learning approach. Plus, performing active learning for a local model that has a training size larger than 12\% (1200) is costly. The improvement with active learning is as high as 22\% when the initial training set is 5.6\% of the 10k adversary's set. 

This improvement with active learning is significant because we can increase the success rate on average from 35\% to 50\% when the number of samples for training the local model is 280 (or 2.8\%). It means our attack traffic flow will bypass once in every two attacks launched. For instance, for an infiltration attack, it is acceptable not to pass the NIDS on the first try. But, at the second attempt, it will most likely bypass the system. To conclude, the adversary can amplify the attack algorithm's success rate by as high as 22\% in the restricted-blackbox threat model.


\subsection{Comparison with Previous GAN-based Attack Algorithms}

\begin{table}[h]
\centering
\caption{Attack comparisons in the blackbox threat model where a local model is trained with 60k samples  \label{compare_gans1}}
\scalebox{0.9}{
{\begin{tabular}{cccc} 
\hline
Constraints & ADVGAN & IDSGAN & NIDSGAN\\ 
\hline
No constraints & 1.00 & 1.00 & 1.00  \\
All constraints enforced & 0.00 & 0.19 & 0.96  \\
\hline
\end{tabular}}{}}
\end{table}

Without constraints, as shown in Table~\ref{compare_gans1}, our NIDSGAN can get nearly 100\% success rates in the blackbox threat model, which simulates the threat model specified in the IDSGAN paper~\cite{idsgan}. The reason for their high success rates in the blackbox threat model is that IDSGAN uses half of the training set of the NSL-KDD dataset (about 60k samples) to train their GANs, which is not much different from a whitebox threat model. Certainly, in many realistic environments, the adversary is unlikely to be able to acquire 60k samples from a targeted NIDS. As shown in the whitebox results before, it is easy to achieve near 100\% success rates even with constraints were to have that kind of access. However, when the domain constraints are enforced to the adversarial traffic flows, the ADVGAN, IDSGAN, and NIDSGAN achieve 0\%, 19.2\%, and 96\% success rates, respectively. The ADVGAN achieves 0\% success rate because it was developed for the image domain, therefore, it does not have a functionality to take the constraints into account. The IDSGAN achieves only 19.2\% success rate when the valid domain constraints are enforced into the adversarial traffic flows. 

\begin{table}[h]
\centering
\caption{Attack comparisons in the restricted-blackbox threat model by success rate \label{compare_gans2}}
\scalebox{0.9}{
{\begin{tabular}{cccc} 
\hline
Dataset & ADVGAN & IDSGAN & NIDSGAN\\ 
\hline
No constraints & 1.00 & 1.00 & 1.00  \\
All constraints enforced & 0.00 & 0.19 & 0.75  \\
\hline
\end{tabular}}{}}
\end{table}

Without constraints, as shown in Table~\ref{compare_gans2}, all algorithms achieve nearly 100\% success rates. However, the success rates of the adversarial traffic flows generated with constraints are significantly lower than those of the previous blackbox threat model. The reason is that the algorithms can not gain much information from the target NIDS with only 740 samples. The reason that our algorithm achieves much better than IDSGAN is that the constraints of our algorithm are significantly tighter and realistic than these of the IDSGAN. Also, we apply active learning techniques to amplify our success rates.  Lastly, we show that active learning can reduce the local model's training size. By using three times smaller local model's training size, we can achieve comparable results when the local model's training size is greater than 2400 as shown in Table. \ref{tab:lessTrainingSize}.

\subsection{Reducing the Local Model's Training Samples}
Finding a large number of training samples for the local model can be very challenging for the adversary in blackbox setting. In this section, we show that active learning can help to reduce the local model's training size by a factor of two (or three) while still achieving comparable results. This shows that even with low resources and capabilities, the adversary can manage to achieve high success rates with active learning in restricted threat models.

\vspace{0.5em}
\begin{table}[h]
\centering
\caption{Reducing the local model's training size with active learning. \label{tab:lessTrainingSize}}
\scalebox{0.77}
{\begin{tabular}{rrcrc} 
\hline
\cline{1-5} 
GAN framework  & \multicolumn{2}{c}{\textbf{Original}} & \multicolumn{2}{c}{\textbf{Active Learning}}\\
\cline{1-5} 
Number of queries  & Local Set & Success Rate & Local Set & Success Rate\\ 
\hline
1.4\% \ppp & $140\,(2x)$ & \textbf{0.19} & $70\,(x) $& \textbf{0.24}\\ 
2.8\%  \ppp & $280\,(2x)$ & \textbf{0.39} &$ 140\,(x)$ & \textbf{0.39}\\
5.6\%  \ppp& $560\,(2x)$ & \textbf{0.44} & $280\,(x) $& \textbf{0.50}\\
7.4\%  \ppp& $740\, (2x)$ & \textbf{0.60} &$ 370\,(x) $ & \textbf{0.51}\\
12.0\%  \ppp& $1200\,(3x)$ & \textbf{0.75} & $400\,(x)$  & \textbf{0.56}\\ 
24.0\% \ppp& $2400\,(3x)$ & \textbf{0.80} & $800\,(x)$  & \textbf{0.75}\\
36.0\% \ppp& $3600\,(3x)$ & \textbf{0.81} & $1200\,(x)$  & \textbf{0.80}\\
67.0\% \ppp& $6700\,(3x)$ & \textbf{0.89} & $2333\,(x) $ & \textbf{0.87}\\
100.0\% \ppp& $10000\,(3x)$ & \textbf{0.96} & $3333\,(x) $ & \textbf{0.93}\\ 
\hline
\multicolumn{5}{c}{NSL-KDD evaluated with IdsNet}\\
\end{tabular}}{}
\end{table}

\subsection{Transfer-based Attack}

\begin{table}
\centering
\caption{Success rate of the NIDSGAN against classical ML algorithms-based NIDS with $\| \Delta x \|_2$ less than 0.3. \label{tab:transfer_attacks}}
\scalebox{0.9}
{\begin{tabular}{ccc} 
\hline
NIDS based on & \multicolumn{2}{c}{\textbf{Adversarial Success Rate}}\\ 
\cline{2-3} 
ML models & CICIDS-2017 & NSL-KDD  \\ 
\hline
Decision Tree & 0.66 & 0.67 \\ 
Support Vector Machine & 0.75 & 0.84   \\
K Nearest Neighbor & 0.66 & 0.44   \\
Logistic Regression & 0.73 & 0.46  \\
\hline
\multicolumn{3}{c}{AlertNet and DeepNet}\\
\end{tabular}}{}
\end{table}

We evaluate our attack algorithm against NIDS models based on classical ML algorithms: Decision tree (DT), support vector machine (SVM), k-nearest neighbors (KNN), and logistic regression (LR) in whitebox setting. As shown in Table.~\ref{tab:transfer_attacks}, the adversarial network traffic flows generated to bypass AlertNet and DeepNet NIDS models by our algorithm can bypass (or transfer) the classical ML algorithm-based NIDS models with 70\% and 61\% of success rates on average in CICIDS-2017 and NSL-KDD cases, respectively.

%% file: 6.Discussion.tex
\vspace{-0.2cm}
\section{Discussion}

One future direction is to explore how little data is necessary to launch attacks.  For example, looking at the current data sets we may focus on how to reduce the local model's training size from 740 to something under 100 for CICIDS-2017 and under 50 for NSL-KDD and still achieve the same high success rates. Techniques such as active learning may provide a path to success.  

Conversely, the active learning approach used in this work comes with other challenges. Here, there is a resource and adversary exposure cost resulting from traffic flows being launched in the networks and analyzing the subsequent reactions of the NIDS.  Here there is a trade-off between the exposure of the adversary and the success rate. Finding the balance between these two metrics becomes a challenge for the adversary. One creative solution is to perform a slow attack by launching attack traffic flows periodically within an extended period of time as long as weeks or months and analyzing the NIDS' responses. That way, the adversary can gain more information about the NIDS without a high risk of getting exposed. However, the longer the local model's training it is, the larger the chance it will be in the decision behavior of the NIDS models may change (e.g., via dynamic training). 

In future work, we will investigate countermeasures to the ML-aided attack algorithms. In the image domain, many defenses are proposed in the literature, such as adversarial training, denoising, distillation, and certification. However, these defenses may not be effective in real NIDS environments for the following three reasons. First, all of these defenses are beaten by stronger follow-up attacks. Second, any defense, excluding the certified defenses can be effective against only one type of attack. Third, these defenses are effective under the assumption that the perturbation threshold is very small and the norm type for the perturbation is given. In addition, we do not know how these defenses will generalize well against unknown attacks, which makes creating an effective defense against adversarial examples difficult. 

We posit that one possible defense could incorporate both ML and rule-based techniques for the defense by looking at similar features of similar network traffic flows. The intuition here is that the attacker could probing the NIDS by launching similar attacks (with similar features) to gain information about NIDS' decision behavior, i.e., detect that reconnaissance is occurring. In that case, the defender could cluster the network traffic flows and carefully compare their features. If the Euclidean distance between the features is extremely small, that signal that the attacker is trying to enforce network constraints for the attack traffic flows. If one of the network flows of the cluster is detected as an attack flow by the NIDS, the chance that the remaining network flows are attack flows is much higher. To conclude, for this defense, we try to capitalize on an important weakness of the attack: the need for a number of queries for both probing the NIDS and for enforcing the domain constraints.

%% file: 7.Conclusion.tex
\vspace{-0.2cm}
\section{Conclusion}
\label{sec:conclusion}
In this work, we developed an attack algorithm NIDSGAN that generates realistic adversarial network traffic flows that bypass DNN-based NIDS models in practical threat models. We show that adversarial traffic flows generated by our attack can bypass NIDS models with 99\%, 85\%, and 70\% success rates in whitebox, blackbox, and restricted-blackbox threat models, respectively. We also show that we can amplify the success rate or reduce the local model's training samples with active learning techniques. We also show that adversarial traffic flows generated by our algorithm can evade classical ML algorithms-based NIDS. 
In future work, we will investigate how to defend these ML-based NIDS against ML-aided attack algorithms.

%% file: appendix.tex
\section{Appendix}
\subsection{Constraint Dictates the Adversarial Success Rates}
 It is easier to evade the models trained with the CICIDS-2017 dataset than the models trained with NSL-KDD dataset. The effect of the domain constraints on the success rate of the attack algorithm is directly related to the complexity and structure of the datasets. We found that our GAN-based attack algorithms can achieve as high as 100\% success rate with local model's training size of only 740 as shown in Figure \ref{compare_gans1}. However, as the constraints are applied to the adversarial traffic flows, the success rates of the algorithms drop to 0\%, 19\%, and 75\% for AdvGAN, IDSGAN, and NIDSGAN, respectively.

\subsection{Imitating Decision Boundary of the Target Models with a Local Model}
\begin{figure}[h]
    \centering
    \includegraphics[width=0.48\textwidth]{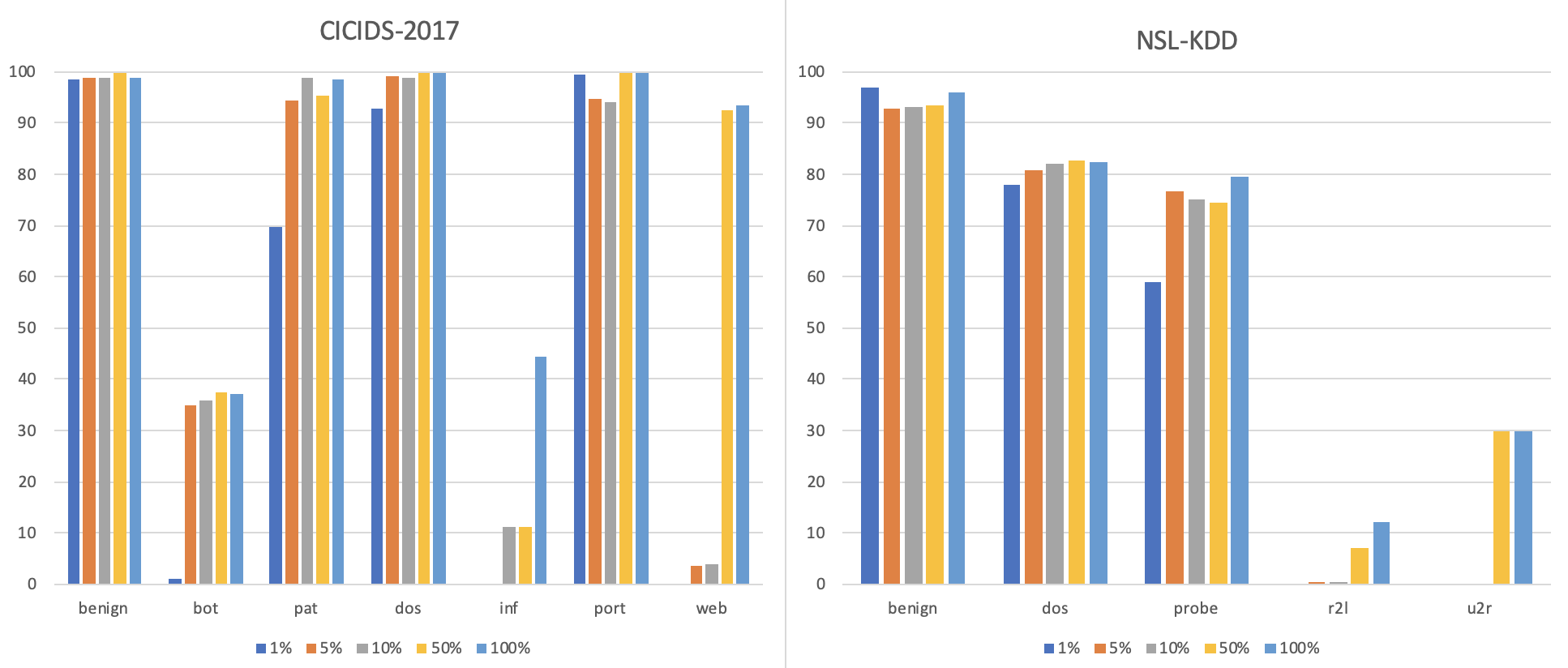}
    \caption{Prediction accuracy of local models with various training sizes}
    \label{fig:compare_class}
    \end{figure}

With a small number of local model's samples, the adversary can approximate the decision behavior of the target models really well with given datasets. With the local model's training size as small as 1\% of the whole training datasets, the adversary can build a local NIDS model that has comparable prediction accuracy to the target model's prediction accuracy as shown in Figure~\ref{fig:compare_class}. 

\subsection{Choice of $l^2$ Norm on Loss}

\begin{table}[htbp]
\centering
\caption{Average perturbation percentage for a single feature based on $\| \Delta x \|_2$ less than 0.3 \label{tab:pertMeasure}}
\scalebox{0.9}
{\begin{tabular}{c|ccccc} 
\hline
number of & \multicolumn{5}{c}{$\Delta x_1$}\\ 
\cline{2-6} 
features perturbed & 0.10 & 0.15 & 0.20 & 0.25 & 0.30\\ 
\hline
30  & 0.052 & 0.04 & 0.04 & 0.03 & 0.0\\
20  & 0.064 & 0.05 & 0.05 & 0.03 & 0.0\\ 
10  & 0.094  & 0.08 & 0.07 & 0.05 & 0.0\\ 
5   & 0.14 & 0.12 & 0.11 & 0.08 & 0.0\\ 
2   & 0.28  & 0.25 & 0.22 & 0.16 & 0.0\\ 
1   & 0.30    & 0.30   & 0.30    & 0.30 & 0.30\\
\hline
\end{tabular}}{}
\end{table}

We pick $l^2$ norm to bound our perturbations because (1) the exceedingly high magnitude of modifications onto a feature would easily break the both the validity of constraints and functionality of attack semantics; (2) large $l^2$-measured perturbations render higher detectability against statistics-based NIDS in general. For these two reasons, $l^2$ norm well suits for our attack algorithm's objective as well as helps to satisfy the constraints of the attacker. 

As shown in Table. \ref{tab:pertMeasure}, when $\| \Delta x \|_2$ perturbation is 0.3, the maximum number of features we can perturb by 0.3 is only one. When $\| \Delta x \|_\infty$ perturbation is 0.3, there is a possibility that the algorithms change all features by 0.3. Here, we show a rough estimate of how much perturbation can be applied to attack traffic's features by varying the number of features perturbed when $\| \Delta x \|_2$ is 0.3. For instance, 10 features out of 123 features are perturbed, the average perturbation level of 9 features are 0.094, 0.08, 0.07, 0.05, 0.0 when fixing $| \Delta x_1 |$ (the one of the 10 perturbed features) values at 0.1, 0.15, 0.2, 0.25, 0.3, respectively. (1) $l^\infty$ vs $l^2$: $l^2$ norm enforces to make small changes to many features while $l^\infty$ norm allows to make large changes to many features. For instance, $l^\infty$=0.3 allows to perturb all of the features by 0.3 while $l^2$=0.3 allows to perturb only one feature by 0.3 or perturb 30 features by 0.055 on average. (2) $l^0$ vs $l^2$: $l^0$ norm minimizes the number of features to perturb. This can have two downsides: 1) it perturbs the features by any amount, which can make the attack easily detectable by finding an outlier for a particular perturbed feature. 2) when the adversary perturbs a small number of features less than five, it could give up the characteristics of the attack traffic that defines its semantics and adopt the behavior that defines the characteristics benign traffic. $l^\infty$ and $l^0$ norms could result in making large changes to features, which is a big risk for the adversary.

For the whitebox cases, we also notice that the difference between constrained and unconstrained cases are largest when the perturbation level is small. For the CICIDS-2017 dataset, the success rate of the constrained case is significantly lower than that of the unconstrained case when $\| \Delta \vec{x}\|_2$ is less than 0.1. As shown in Figure~\ref{fig:final}, the largest discrepancy between constrained and unconstrained cases occurs when the $\|\Delta \vec{x}\|_2$ perturbation is 0.01, which is approximately 50\% of success rate. The difference slowly shrinks as the perturbation increases to 0.1. In the NSL-KDD dataset, the difference between the success rates of the two cases is largest when the perturbation is 0.12. The difference shrinks as the perturbation increases and becomes zero when the perturbation is 0.275. 

\subsection{Additional Information about the Datasets and the Training}
We observed that the percentages of benign traffic flows in CICIDS-2017 and NSL-KDD are 83\% and 52\%, respectively. The territory of benign traffic flows in the feature space is significantly larger than that of any other type of traffic flow, which is more dominant in the CICIDS-2017 case. Benign dominant highly imbalanced dataset could benefit the adversary. To analyze that, we balanced the datasets with oversampling techniques such as SMOTE and ADASYN; however, we did not notice much change in the NIDS' prediction accuracy as well as the effectiveness of our algorithm.  

\begin{table}[htbp]

\caption{Breakdown of different subclasses of each attack of NSL-KDD dataset \label{tab:nsl_kdd_additional}}
\scalebox{0.6}
{\begin{tabular}{|c|c|c|c|c|c|} 
\hline
Classes & \texttt{Benign} & \texttt{DoS} & \texttt{Probe} & \texttt{U2R} & \texttt{R2L}\\ 
\hline
Sub-Classes & benign   & apache2  & ipsweep & buffer\_overflow & ftp\_write\\
            &           & back     & mscan & loadmodule        & guess\_passwd\\
 & & land     & nmap & perl               & httptunnel\\
 & & neptune  & portsweep & ps            & imap\\
 &  & mailbomb & saint & rootkit           & multihop\\
 &  & pod      & satan & sqlattack         & named\\
 & & processtable &  & xterm              & phf\\
 &  & smurf    &  &                        & sendmail\\
 &  & teardrop &  &                        & Snmpgetattack\\
 &  & udpstorm &  &                        & spy\\
&   & worm     &  &                        & snmpguess\\
&   &          &  &                        & warezclient\\
&   &          &  &                        & warezmaster\\
&   &          &  &                        & xlock\\
&   &          &  &                        & xsnoop\\
\hline
Size   & 52.1\%  & 36.1\%  & 9.4\% & 0.1\% & 2.3\% \\
\hline
\end{tabular}}{}
\end{table}

\begin{table}[htbp]

\caption{Breakdown of different subclasses of each attack of CICIDS-2017 dataset \label{tab:CICIDS_additional}}
\scalebox{0.48}
{\begin{tabular}{|c|c|c|c|c|c|c|c|} 
\cline{1-8} 
Classes & \texttt{Benign} & \texttt{Botnet} & \texttt{Patator} & \texttt{DoS/DDoS} & \texttt{Infiltration} & \texttt{PortScan} & \texttt{Web Attacks}\\ 

\hline
Sub-Classes  & Benign  & Bot    & FTP-Patator       & DDoS                  & Infiltration  & PortScan          & Brute Force\\
             &         &        & SSH-Patator       & DoS                   &               &                   & Sql Injection\\
             &          &       &                   & GoldenEye             &               &                   & XSS\\
             &          &       &                   & DoS Hulk              &               &                   & \\
             &          &       &                   & DoS Slow-httptest     &               &                   & \\
             &          &       &                   & DoS slowloris         &               &                   & \\
             &          &       &                   & Heartbleed            &               &                   & \\
\hline
Size & 83.3\% & 0.08\% & 0.5\%& 10.4\% & 0.001\% & 5.6\% & 0.07\% \\
\hline
\end{tabular}}{}
\end{table}

\subsection{Metrics for measuring the prediction accuracy}
Precision is defined as the ratio of the number of outcomes correctly classified as positive to the total number of outcomes classified as positive (the sum of true positives and false positives).  
\begin{equation}
 \label{metrics}
\textrm{Precision} = \frac{\textrm{TP}}{\textrm{TP}+\textrm{FP}}.
\end{equation}
Recall is defined as the number of outcomes correctly classified as positive to the total number of positive outcomes (those that are true positive added to those incorrectly labeled false negatives).  
\begin{equation}
\textrm{ Recall} = \frac{\textrm{TP}}{\textrm{TP}+\textrm{FN}}.
\end{equation}
$F_1$-score is the harmonic mean of precision and recall.
\begin{equation}
\textrm{$F_1$-score} =  2 \, \frac{\textrm{Precision} \cdot  \textrm{Recall}}{ \textrm{Precision} + \textrm{Recall}}.
\end{equation}



 %
    
\begin{table}[t]
\centering
\caption{Realization of feature-space manipulations in live traffic streams} \label{tab:threat_model}
\scalebox{0.8}{
\begin{tabular}{|p{50mm} | p{6mm} | p{6mm} | p{6mm} | p{13mm}|} 
\hline
\textbf{Features} & \textbf{Fwd} & \textbf{Bwd} & \textbf{Both} & \textbf{Operation}\\ 
\hline
Total Duration  
& \makecell{\xmark } 
& \makecell{\xmark } 
& \makecell{\cmark } 
& \makecell{Rearrange} \\
\hline

\hline
Total Packets 
& \makecell{\cmark } 
& \makecell{\cmark } 
& \makecell{\xmark } 
& \makecell{Split} \\
\hline

\hline
Total Length of Packets  
& \makecell{\cmark } 
& \makecell{\cmark } 
& \makecell{\xmark } 
& \makecell{Inject} \\
\hline

\hline
{Pkt Len Min/Max/Mean/Stddev} 
& \makecell{\cmark } 
& \makecell{\cmark } 
& \makecell{\cmark } 
& \makecell{Inject} \\
\hline

\hline
{IAT Min/Max/Mean/Stddev}  
& \makecell{\cmark } 
& \makecell{\cmark } 
& \makecell{\cmark } 
& \makecell{Rearrange} \\
\hline

\hline
Bytes/s 
& \makecell{\xmark } 
& \makecell{\xmark } 
& \makecell{\cmark } 
& \makecell{Rearrange} \\
\hline

\hline
Pkts/s 
& \makecell{\cmark } 
& \makecell{\cmark } 
& \makecell{\cmark } 
& \makecell{Rearrange} \\
\hline

\hline
PSH/URG Flags
& \makecell{\cmark } 
& \makecell{\cmark } 
& \makecell{\cmark } 
& \makecell{Inject} \\
\hline

\hline
FIN/SYN/RST/ACK/CWE/ECE Flags
& \makecell{\xmark } 
& \makecell{\xmark } 
& \makecell{\cmark } 
& \makecell{Inject} \\
\hline

\hline
Total Length of Headers  
& \makecell{\cmark } 
& \makecell{\cmark } 
& \makecell{\xmark } 
& \makecell{Inject} \\
\hline

\hline
Down/Up Ratio
& \makecell{N/A } 
& \makecell{N/A } 
& \makecell{\cmark } 
& \makecell{Inject} \\
\hline

\hline
Avg Bytes/Bulk  
& \makecell{\cmark } 
& \makecell{\cmark } 
& \makecell{\xmark } 
& \makecell{Split} \\
\hline

\hline
Avg Packets/Bulk
& \makecell{\cmark } 
& \makecell{\cmark } 
& \makecell{\xmark } 
& \makecell{Inject} \\
\hline

\hline
Avg Bulk Rate
& \makecell{\cmark } 
& \makecell{\cmark } 
& \makecell{\xmark } 
& \makecell{Inject} \\
\hline

\hline
Initial Window Bytes
& \makecell{\cmark } 
& \makecell{\cmark } 
& \makecell{N/A } 
& \makecell{Inject} \\
\hline

\hline
Packets w/ payload >= 1
& \makecell{\cmark } 
& \makecell{\xmark } 
& \makecell{\xmark } 
& \makecell{Split} \\
\hline

\hline
Min. Packet Header Size 
& \makecell{\xmark } 
& \makecell{\cmark } 
& \makecell{\xmark } 
& \makecell{Inject} \\
\hline

\hline
Active Time Min/Max/Mean/Stddev  
& \makecell{\xmark } 
& \makecell{\xmark } 
& \makecell{\cmark } 
& \makecell{Rearrange} \\
\hline

\hline
Idle Time Min/Max/Mean/Stddev 
& \makecell{\xmark } 
& \makecell{\xmark } 
& \makecell{\cmark } 
& \makecell{Rearrange} \\
\hline
\end{tabular}

}{}
    {\raggedright 
    \textbf{Fwd}, \textbf{Bwd}, and \textbf{Both} indicate if the feature is calculated for the specified directions. The \textbf{Operation} column lists the deployable packet manipulation techniques in real traffic streams for perturbing the corresponding feature value. Our deployable manipulations include three concrete strategies: 1) \textbf{\textrm{Inject}} -- inserting non-functional packets with desired lengths, transmission times, and flag combinations; 2) \textbf{\textrm{Rearrange}} -- adjusting the time intervals between consecutive packets as per the perturbation needs; and 3) \textbf{\textrm{Split}} -- increasing the number of packets by dividing the payload in one packet into more packets with smaller payload segments~\cite{hashemi2019towards} without disrupting the intended packet semantics. \par}
    \label{tab:mapback}
\end{table}